\documentclass[11pt]{article}
\usepackage{amssymb}

\usepackage{graphicx}
\usepackage{amsmath}
\usepackage{makeidx}
\usepackage{indentfirst}

\newcounter{resultnum}[section]

\newcounter{conclusionnum}[section]

\newcounter{conditionnum}[section]

\newcounter{conjecturenum}[section]

\newcounter{examplenum}[section]

\newcounter{exercisenum}[section]

\newcounter{lemmanum}[section]

\newcounter{notationnum}[section]

\newcounter{theoremnum}[section]

\newcounter{definitionnum}[section]

\newcounter{corollarynum}[section]

\newcounter{remarknum}[section]

\newcounter{propositionnum}[section]

\newcounter{acknowledgementnum}[section]

\newcounter{algorithmnum}[section]

\newcounter{axiomnum}[section]

\newcounter{casenum}[section]

\newcounter{claimnum}[section]

\newcounter{summarynum}[section]

\newcounter{problemnum}[section]

\begin{document}

\title{Modified Dispersion Relations in\\
Ho\v rava--Lifshitz Gravity and Finsler Brane Models}
\date{December 16, 2011}
\author{\textbf{Sergiu I. Vacaru}
 \thanks{sergiu.vacaru@aei.mpg.de, sergiu.vacaru@uaic.ro, Sergiu.Vacaru@gmail.com} \\
%EndAName
\textsl{\small University "Al. I. Cuza" Ia\c si, Science Department,} \\
\textsl{\small 54 Lascar Catargi street, Ia\c si, Romania, 700107 } }
\maketitle

\begin{abstract}
We study possible links between quantum gravity phenomenology encoding Lorentz violations as nonlinear dispersions, the Einstein--Finsler gravity models, EFG, and nonholonomic (non--integrable) deformations to Ho\v
rava--Lifshitz, HL, and/or Einstein's general relativity, GR, theories. EFG and its scaling anisotropic versions formulated as Ho\v rava--Finsler models, HF, are constructed as covariant metric compatible theories on (co) tangent bundle to Lorentz manifolds and respective anisotropic deformations. Such theories are integrable in general form and can be quantized following standard methods of deformation quantization, A--brane formalism and/or (perturbatively) as a nonholonomic gauge like model with bi--connection
structure. There are natural warping/trapping mechanisms, defined by the maximal velocity of light and locally anisotropic gravitational interactions in a (pseudo) Finsler bulk spacetime, to four dimensional (pseudo) Riemannian spacetimes. In this approach, the HL theory and scenarios of
recovering GR at large distances are generated by imposing nonholonomic constraints on the dynamics of HF, or EFG, fields.


\end{abstract}

%\tableofcontents

\section{Introduction}

There is a considerable recent interest in two directions of classical and
quantum gravity and possible implications in cosmology and astrophysics:\
The first one is related to gravity models with anisotropic scaling between
space and time at short distances which is usually referred to as the Ho\v{r}%
ava--Lifshitz, HL, theory \cite{horava1,horava2,horava3}. Such theories with
generic anisotropy are usually non--relativistic and ultra--violet complete;
the local Lorentz invariance is violated/brocken (LV) at short distances but
constructed to reduce to the general relativity (GR) theory in the infrared
limit.\footnote{%
The problem of reduction is still an open issue: the HL theory with global
Hamiltonian does not reproduce GR in the infrared domain \cite{kobakh}.
There are necessary certain further modifications of the theory because both
projectable and non--projectable versions of the HL models seem to contain
certain inconsistency \cite{blas,odintsov,carloni}.}  One of the main features of this class
of theories is that they can be elaborated in a "power--counting"
renormalizable form (unlike Einstein gravity) at least if the so called
detailed balance condition is respected. This can be understood, for
instance, as a result of stochastic quantization in relation to topological
massive gravity \cite{orlando1,orlando2}. The second direction consists from
a series of models related to quantum gravity (QG) phenomenology also
including LV effects and general relativistic and non--relativistic
constructions with extra dimensions, generalized symmetries, and
compactification or trapping scenarios etc (see, for instance, reviews \cite%
{kost4,xiao,liberati,carroll,burgess,barcello}).

The above mentioned classes of gravitational theories are characterized, in
general, by LV and respective modified dispersion relations (MDR), local
and/or global anisotro\-pi\-es, nonhomogeneous and, for certain models, they
are defined by nonholonomic (non--integrable) constraints on the dynamics of
gravitational and matter fields. For instance, the implications of
violations of Poincar\' e symmetry for kinematic conditions and MDR at the
"threshold" for some particle--creation interactions in HL--type theories
are studied in \cite{mercati}. One of the most important aspects to be
understood is the way when such theories can be constructed in a general
geometric form and to analyze possible applications.

During last decade, there have been published some series of works relating
quantum phenomenology and, for instance, anisotropy and dark energy/matter
problems to Finsler gravity models with LV, MDR and locally anisotropic
spacetime configurations, see explicit constructions and references in \cite%
{mavromatos,mignemi,girelli,gibbons,sindoni,
lammer,stavrinos,stavr0,visser1,visser2,yang,calcagni,schuller}. There is a
study of possible links between anisotropic--scaling scenarios and Finsler
spacetimes \cite{sindoni1}. A surprising conclusion which can be drawn from
such approaches is that we have to include certain Finsler type physical
objects into various schemes of quantization of gravity and apply
corresponding geometric methods in order to elaborate in a self--consistent
form relativistic and non--relativistic models of QG.

There were proposed different ideas and elaborated explicit theoretical
constructions related to Finsler geometry, generalizations and applications
in modern physics (for particle and mathematical physics researches, we cite
Refs. \cite{vrev1,vcosm,vbrane,vsgg,vcrit,ma}\footnote{%
there are thousands of papers and tens of monographs on Finsler geometry and
applications - it is not possible to summarize in this paper and discuss all
such ''standard'' and ''nonstandard'' theories in relation to modern gravity
and analogous mechanical models}). For instance, Finsler gravity models can
be derived in low energy limits from string gravity theories \cite%
{vstr1,vstr2} and brane gravity \cite{vsingl1,vsingl2,vbrane} and induced by
noncommutative generalizations of Einstein gravity \cite{vnc1,vnc2,vnc3}.
Various classes of commutative and noncommutative Finsler type geometries
and gravity theories are induced via nonholonomically constrained Ricci
flows of (pseudo) Riemannian metrics \cite{vrf1,vrf2,vrf3,vrf4}. Finsler
variables can be introduced in GR and extra dimension generalizations which
allows us to formulate geometric methods of constructing exact (and vary
general classes of) solutions in different gravity theories \cite%
{vex1,vex2,vex3,vsgg,vnc2}. Re--writing the Einstein equations in the
so--called almost K\"{a}hler -- Finsler variables, it was possible to apply
rigorous methods of deformation and A--brane quantization and nonholonomic
gauge methods in order to elaborate quantum models of Einstein gravity and
Lagrange--Finsler--Hamilton generalizations \cite%
{vqgr1,vqgr2,vqgr3,vqgr4,vqgr5,vqgr6}.

It is our purpose to elaborate a modification of HL and GR theories which
will include MDR defined, in general, for tangent bundles to Einstein
spacetimes and nonholonomic/ anisotropic modifications. Naturally such
constructions can be performed in the framework of Finsler geometry and
generalizations. We consider such an approach to be motivated because any
type of nonlinear dispersion relations are canonically related to a Finsler
generation function (usual constructions in special and general relativity
theories are contained as particular (quadratic) cases). In this sense, QG
models of HL and/or other types should be more realistically elaborated in
terms of (pseudo) Finsler fundamental geometric objects; this seems to give
a more realistic quantum theory the existing quantum versions of (pseudo)
Riemannian geometry.

The paper is organized as follows. In section \ref{sec2}, we provide a brief
summary of HL and GR theories and consider possible MDR for scaling
anisotropies. We show that certain class of fundamental Finsler functions
can be derived from HL theory and various types of QG models with MDR. We
formulate the HF gravity as a theory generalizing HL models on Finsler
spaces in section \ref{sec3}. Trapping mechanism for Finsler branes
resulting in HL gravity (and for corresponding nonholonomic constraints, in
GR theory) are studied in section \ref{sec4}. Finally, conclusions are
provided in section \ref{sec5}. In Appendix, we summarize some technical
details on diagonal solutions in HF gravity.

\section{Finsler Geometry Induced by MDR in HL Gravity}

\label{sec2} The goal of this section is to show how fundamental Finsler
geometric objects are induced from some general MDR and, in particular, for
HL gravity: We outline in brief the HL gravity theory, analyze possible MDR
and show how the fundamental Finsler generating function can be associated
to such anisotropic configurations and nonlinear dispersions.

\subsection{Preliminaries on the HL and GR theories}

In standard form, the dynamical variables of HL gravity are the lapse
function, $N,$ the shift function, $N^{\widehat{i}},$ and the spacelike
metric, $g_{\widehat{i}\widehat{j}},$ in terms of which the metric is
written as the Arnowitt--Deser--Misner, ADM, (1+3) splitting,%
\begin{equation}
ds^{2}=g_{ij}dx^{i}dx^{j}=-N^{2}dt^{2}+g_{\widehat{i}\widehat{j}}(dx^{%
\widehat{i}}+N^{\widehat{i}}dt)(dx^{\widehat{j}}+N^{\widehat{j}}dt).
\label{adm}
\end{equation}%
The above metric $g_{ij}=(N^{2},g_{\widehat{i}\widehat{j}})$ \ (we shall
write in brief, $hg=(N^{2},\widehat{g}))$ is supposed to be invariant under
the foliation--preserving diffeomorphisms of the HL theory, $t^{\prime
}=t^{\prime }(t)$ and $x^{\widehat{i}^{\prime }}=x^{\widehat{i}}(t,x^{%
\widehat{k}}),$ where indices $i,i^{\prime },j,j^{\prime },...=1,2,3,4,$ for
$x^{i}=(x^{1}=t,x^{\widehat{i}})$ and $\widehat{i},\widehat{i}^{\prime },%
\widehat{j},\widehat{j}^{\prime },...=2,3,4.$\footnote{%
We have to elaborate a new system of notations which will be compatible with
3+1 splitting for ADM formalism and 4+4, or 2+2/3+2 / 4+3 nonholonomic
splitting used in Finsler geometry, see details in \cite{vrev1}.} The theory
is invariant under the anisotropic scaling symmetry%
\begin{equation}
t\rightarrow \mathit{l}^{z}t,\ x^{\widehat{i}}\rightarrow \mathit{l}x^{%
\widehat{i}},\mbox{ when for }z=3,\ N\rightarrow \mathit{l}^{-2}N,N^{%
\widehat{i}}\rightarrow \mathit{l}^{-2}N^{\widehat{i}},\ g_{\widehat{i}%
\widehat{j}}\rightarrow g_{\widehat{i}\widehat{j}}  \label{scalin}
\end{equation}%
(to elaborate a power--counting renormalizable theory of gravity in four
dimensions, 4--d, is considered $z=3).$ The projectability condition
requires a homogeneous lapse function $N=N(t)$ but admits general shift and
3--d metric, i.e. $N^{\widehat{i}}(x^{k})=N^{\widehat{i}}(t,x^{\widehat{k}})$
and $\ g_{\widehat{i}\widehat{j}}(x^{k})\rightarrow g_{\widehat{i}\widehat{j}%
}(t,x^{\widehat{k}}).$\footnote{%
It is possible to consider a general nonhomogeneous lapse function but this
may result in problems when attempting to quantize the model, see \cite%
{horava2,li}.}

The action for HL gravity is postulated as a sum of ''kinetic'', $\ _{K}S,$
and ''potential'' part, $\ _{V}S,$%
\begin{equation}
\ ^{HL}S=\ _{K}S+\ _{V}S,  \label{action}
\end{equation}%
where
\begin{eqnarray*}
\ _{K}S &=&\frac{2}{\kappa ^{2}}\int dtd^{3}x\sqrt{|\widehat{g}|}N\left( K_{%
\widehat{i}\widehat{j}}K^{\widehat{i}\widehat{j}}-\lambda K^{2}\right) \\
\ _{V}S &=&\int dtd^{3}x\sqrt{|\widehat{g}|}N[\frac{\kappa ^{2}\mu }{2\varpi
^{2}}\epsilon ^{\widehat{i}\widehat{j}\widehat{k}}R_{\widehat{i}\widehat{l}%
}\nabla _{\widehat{j}}R_{\ \widehat{k}}^{\widehat{l}}-\frac{\kappa ^{2}\mu }{%
8}R_{\widehat{i}\widehat{j}}R^{\widehat{i}\widehat{j}} \\
&&+\frac{\kappa ^{2}\mu }{8(1-3\lambda )}\left( \frac{1-4\lambda }{4}%
R^{2}+\Lambda R-3\Lambda ^{2}\right) -\frac{\kappa ^{2}}{2\varpi ^{2}}C_{%
\widehat{i}\widehat{j}}C^{\widehat{i}\widehat{j}}],
\end{eqnarray*}%
for some constants $\kappa ,\mu ,\varpi ,\Lambda $ and a dynamical constant $%
\lambda $ running as the energy scale changes. The general covariance in GR
imposes the condition $\lambda =1.$ It is known that the important
variation-interval of $\lambda $ is between $1/3$ (the ultra--violet, UV,
limit) and $1$ (the infra--red, IR, limit). In the above formulas,
\begin{equation*}
K_{\widehat{i}\widehat{j}}=\frac{1}{2N}\left( \frac{\partial g_{\widehat{i}%
\widehat{j}}}{\partial t}-\nabla _{\widehat{i}}N_{\widehat{j}}-\nabla _{%
\widehat{j}}N_{\widehat{i}}\right)
\end{equation*}%
is the extrinsic curvature with $K=g^{\widehat{i}\widehat{j}}K_{\widehat{i}%
\widehat{j}};$ the Cotton tensor is defined
\begin{equation*}
C^{\widehat{i}\widehat{j}}=\frac{\epsilon ^{\widehat{i}\widehat{j}\widehat{k}%
}}{\sqrt{|\widehat{g}|}}\nabla _{\widehat{k}}\left( R_{\ \widehat{l}}^{%
\widehat{j}}-\frac{1}{4}R\delta _{\ \widehat{l}}^{\widehat{j}}\right) ,
\end{equation*}%
where such geometric objects are constructed for the Levi--Civita connection
$\nabla _{\widehat{k}}$ and $R$ determined by the 3--d spacial metric $g_{%
\widehat{i}\widehat{j}},$ for $\delta _{\ \widehat{l}}^{\widehat{j}}$ being
the Kronecker symbol and $|\widehat{g}|$ computed as the determinant of 3--d
metric. In this work, we shall consider a simple form of theory with field
equations derived from (\ref{action}) (for instance, we can also consider
the ''detailed balance'' condition which reduces the number of terms in the
potential).\footnote{%
The most possible general potential is analyzed in \cite{sotir1,sotir2}; we
shall elaborate more simple constructions which do not change our basic
conclusions on relation of MDR and Finsler geometry.}

In the infrared limit of (\ref{action}) we can obtain the ADM form of the
Einstein--Hilbert action if the speed of light, $c,$ gravitational constant,
$G,$ and cosmological constant, $\ ^{GR}\Lambda ,$ (all in GR) are defined,
respectively,
\begin{equation}
c=\frac{\kappa ^{2}\mu }{4}\sqrt{\frac{\Lambda }{1-3\lambda }},\ 16\pi G=%
\frac{\kappa ^{4}\mu }{8}\sqrt{\frac{\Lambda }{1-3\lambda }}\mbox{ and }\
^{GR}\Lambda =\frac{3\kappa ^{4}\mu ^{2}\Lambda ^{2}}{32(1-3\lambda )}.
\label{grcond}
\end{equation}%
There is also a coefficient before the $R^{2}$ term, $\kappa ^{2}\mu
^{2}=8(1-3\lambda )c^{3}/16\pi G\Lambda .$ The GR theory can be considered
as a ''homogeneous'' and locally isotropic version of HL gravity.

\subsection{MDR in HL gravity}

\label{ssmdr}A stability analysis of HL gravity is performed, for instance,
in Ref. \cite{bogdanos}. The conclusion of that work is that the HL gravity
in original form suffers from instabilities and fine--tuning which cannot be
overcome by simple tricks such an analytic continuation, see also \cite%
{kobakh,blas}. We propose that the HL theory should be extended on
tangent/cotangent bundle (with velocity type coordinates) in order to
include MDR which will put the problem of stability of gravitational field
equations with nonholonomic constraints in a different form.

Let us outline some typical dispersion relations\footnote{%
which can be computed by perturbing the action (\ref{action}) up to second
order of metric preserving the ADM 3+1 foliation preserving formalism around
a flat background} in HL gravity. Under the so--called ''detailed balance''
conditions, there are possible the following variants (with Fourier
transforms of type $\psi (t,x^{\widehat{i}})=\int \frac{d^{3}k}{(2\pi )^{3/2}%
}\psi _{p}(t)e^{ip_{\widehat{i}}x^{\widehat{i}}}).$

\begin{itemize}
\item For scalar perturbations and considering a low--$p$ behavior, we
acquire
\begin{equation*}
\omega ^{2}=-\frac{9\kappa ^{4}\mu ^{2}\Lambda ^{2}}{32(1-3\lambda )^{2}}<0.
\end{equation*}%
Such a MDR induces instabilities at the IR for all values of $\lambda $ and
both sings of $\Lambda .$

\item For high--$p,$ the dispersion relation is
\begin{equation*}
\omega ^{2}=\frac{\kappa ^{4}\mu ^{2}}{16}\left( \frac{1-\lambda }{%
1-3\lambda }\right) ^{2}p^{4}.
\end{equation*}

\item Similar computations can be performed for tensor perturbations,%
\begin{equation*}
\omega ^{2}=c^{2}p^{2}+\frac{\kappa ^{4}\mu ^{2}}{16}p^{4}\pm \frac{\kappa
^{4}\mu }{4\varpi ^{2}}p^{5}+\frac{\kappa ^{4}}{4\varpi ^{4}}p^{6}.
\end{equation*}
\end{itemize}

A perturbative analysis can be extended beyond detailed balance. Such
extended relations can be written (using an additional parameter for the
corresponding contribution to the action):

\begin{itemize}
\item For the UV--behavior of scalar perturbations,%
\begin{equation*}
\omega ^{2}=\frac{\kappa ^{2}(1-\lambda )^{2}}{16(1-3\lambda )^{2}}p^{4}-%
\frac{3\kappa ^{2}(1-\lambda )}{2(1-3\lambda )}\eta p^{6}.
\end{equation*}

\item Finally, we present the formula for tensor perturbations:
\begin{equation*}
\omega ^{2}=c^{2}p^{2}+\frac{\kappa ^{4}\mu ^{2}}{16}p^{4}\pm \frac{\kappa
^{4}\mu }{4\varpi ^{2}}p^{5}+\left( \frac{\kappa ^{4}}{4\varpi ^{4}}-\frac{%
\kappa ^{2}\eta }{2}\right) p^{6}.
\end{equation*}
\end{itemize}

We conclude that HL\ theory with Minkovski background is characterized by
corresponding MDR $\omega (p^{i},\kappa ,\mu ,\Lambda ,\varpi ,c,\lambda
,\eta )$ depending nonlinearly on momentum variables and with critical
behavior (up to instabilities, branching of dispersion relation etc)
determined by the values of the fundamental constants of the theory. The
formulas for nonlinear dispersions presented in this sections are typical
ones which can be derived in various models of HL gravity or alternative
theories (in different approaches, one can be considered only "even powers"
of momenta, parametric deformations etc).

\subsection{Fundamental Finsler functions and the HL theory}

In a more general context, we can perform an analysis of propagation of
light rays in HL and various classes of gravity theories with LV, see
details, for instance, in Refs. \cite{lammer,vcosm,vbrane}. For light rays
propagating on HL\ spacetime, the nonlinear \ dispersion relation\footnote{%
we can consider such a relation in a fixed point \ $x^{k}=x_{(0)}^{k},$ \
when $g_{\widehat{i}\widehat{j}}(x_{0}^{k})=g_{\widehat{i}\widehat{j}}$ and $%
q_{\widehat{i}_{1}\widehat{i}_{2}...\widehat{i}_{2r}}=q_{\widehat{i}_{1}%
\widehat{i}_{2}...\widehat{i}_{2r}}$ $(x_{0}^{k})$} between the frequency $%
\omega $ and the wave vector $k_{i},$ can be written in a general abstract
form
\begin{equation}
\omega ^{2}=c^{2}\left[ g_{\widehat{i}\widehat{j}}k^{\widehat{i}}k^{\widehat{%
j}}\right] ^{2}\left( 1-\frac{1}{r}\frac{q_{\widehat{i}_{1}\widehat{i}_{2}...%
\widehat{i}_{2r}}y^{\widehat{i}_{1}}...y^{\widehat{i}_{2r}}}{\left[ g_{%
\widehat{i}\widehat{j}}y^{\widehat{i}}y^{\widehat{j}}\right] ^{2r}}\right) .
\label{disp}
\end{equation}%
Depending on explicit parametrizations, with $k_{i}\rightarrow p_{i}\sim
y^{a},$ we can include the above dispersion formulas for scalar and tensor
perturbations, or for light propagation, into a formal expression of type (%
\ref{disp}). Such MDR can be derived from very general arguments for a large
class quantum and classical, commutative and noncommutative, gravity and
particle field theories with LV, see \cite%
{kost4,xiao,liberati,dimopoulos,anchor,amelino,carroll,burgess,mercati} (the
coefficients $q_{\widehat{i}_{1}\widehat{i}_{2}...\widehat{i}_{2r}}$ are
computed in explicit form for corresponding models).

In a series of works \cite%
{mavromatos,mavromatos1,mavromatos2,mignemi,girelli,gibbons,sindoni,lammer,stavrinos,stavr0,vcosm,vbrane}%
, there were analyzed various possibilities when MDR (\ref{disp}), or
certain particular forms\footnote{%
for instance, in the very special relativity, with corrections from string
and/or noncommutative dynamics, with Higgs type induced Finsler structures
etc}, can be naturally associated to nonlinear homogeneous quadratic
elements (with $F(x^{i},\beta y^{j})=\beta F(x^{i},y^{j}),$ for any $\beta
>0),$ when
\begin{eqnarray}
ds^{2} &=&F^{2}(x^{i},y^{j})  \notag \\
&\approx &-(cdt)^{2}+g_{\widehat{i}\widehat{j}}(x^{k})y^{\widehat{i}}y^{%
\widehat{j}}\left[ 1+\frac{1}{r}\frac{q_{\widehat{i}_{1}\widehat{i}_{2}...%
\widehat{i}_{2r}}(x^{k})y^{\widehat{i}_{1}}...y^{\widehat{i}_{2r}}}{\left(
g_{\widehat{i}\widehat{j}}(x^{k})y^{\widehat{i}}y^{\widehat{j}}\right) ^{r}}%
\right] +O(q^{2}).  \label{fbm}
\end{eqnarray}%
Such nonlinear metric elements are usually considered in Finsler geometry. A
value $F$ is considered to be a fundamental (generating) Finsler function
usually satisfying the condition that the Hessian
\begin{equation}
\ ^{F}g_{ij}(x^{i},y^{j})=\frac{1}{2}\frac{\partial F^{2}}{\partial
y^{i}\partial y^{j}}  \label{hess}
\end{equation}%
is not degenerate.

For $q_{\widehat{i}_{1}\widehat{i}_{2}...\widehat{i}_{2r}}\rightarrow 0$ and
a corresponding re-definition of frames and coordinates, we can generate
elements of type (\ref{adm}) for GR. The HL theory is with generic
anisotropy and LV characterized by dispersion relations $\omega
(p^{i},\kappa ,\mu ,\Lambda ,\varpi ,c,\lambda ,\eta )$ considered in
section \ref{ssmdr}. Our idea is to extend the Ho\v{r}ava constructions in a
(pseudo) Finsler form on tangent bundles to Lorentz modified manifolds which
will include nonlinear dispersion relations and parametric dependence of
solutions with various stable and nonstable nonlinear properties. Such
Finsler structures are determined naturally from perturbative properties and
light/probing bodies propagations in HL gravity. A Finsler generalization of
HL gravity can be constructed in metric compatible form following principles
very similar to the Einstein and Einstein--Finsler gravity (EFG) \cite%
{vcosm,vbrane,vrev1}, for metric compatible Finsler connections. The
gravitational field equations for such a theory can be integrated in general
form following methods \cite{vex1,vex2,vex3} (with parametric dependence of
solutions which allows us to consider stable and non--stable
configurations). It is also possible to quantize certain classes of Ho\v{r}%
ava--Finsler (HF) gravity \ models following methods of deformation
quantization, A--brane formalism, gauge like methods etc, see \cite%
{vqgr1,vqgr2,vqgr3,vqgr4,vqgr5,vqgr6}.

\section{Ho\v{r}ava--Finsler Gravity}

\label{sec3}In this section, we provide a Finsler generalization of the HL
theory (called the Ho\v rava--Finsler, in brief, HF) which will include as
some ''branch'' configurations respective MDR on tangent bundle $T\mathbf{V}%
, $ where $\mathbf{V}$ is a (pseudo) Riemannian spacetime in GR or its
anisotropic modifications defined by a HL action (\ref{action}).

\subsection{Fundamental geometric objects for HF gravity}

We shall label local coordinates on $T\mathbf{V}$ in the form $u^{\alpha
}=(x^{i},y^{a})$ (in brief $u=(x,y)$), where $x^{i}$ are local coordinates
on $\mathbf{V}$ and $y^{a}$ are fiber (velocity, or momentum type)
coordinates. Indices $\alpha ,\beta ,...$ will run values $1,2,...,8.$

Contrary to the case of (pseudo) Riemannian geometry (which is completely
determined by its metric tensor), a fundamental Finsler metric
(equivalently, generating function) $F^{2}$ (\ref{fbm}) and/or its Hessian $%
\ ^{F}g_{ij}$ (\ref{hess}) do not define completely a geometric/physical
model on $T\mathbf{V.}$ We need certain additional assumptions in order to
construct in a unique form a triple of fundamental geometric objects (a
nonlinear connection, N--connection, structure, a metric structure on the
total space and a linear connection which is adapted to a chosen
N--connection structure, called a distinguished connection, in brief, a
d--connection; in a canonical approach all such objects are induced in a
unique way by fundamental Finsler function $F$), which are necessary for
definition of a physical generalized spacetime/gravitational model using
principles of Einstein--Finsler gravity (EFG) \cite{vcosm,vrev1}.

\subsubsection{N--connections induced by MDR and associated Finsler
generating functions}

A N--connection $\mathbf{N}$ is defined as a Whitney sum%
\begin{equation}
TT\mathbf{V}=hT\mathbf{V}\oplus vT\mathbf{V}.  \label{whitney}
\end{equation}%
With respect to a local coordinate base, it is determined by its
coefficients $\mathbf{N}=\{N_{i}^{a}(x,y)\},$ i.e. $\mathbf{N=}%
N_{i}^{a}dx^{i}\otimes \partial /\partial y^{a}.$\footnote{%
Following our notation conventions \cite{vrev1,vsgg}, we use boldface
symbols for spaces and geometric object on spaces endowed with N--connection
structure. Because there are standard denotations using symbol $N$ both in
ADM model of gravity and in Finsler geometry, we have to use $(N,N^{\widehat{%
i}})$ for lapse and shifting functions and $N_{i}^{a}$ for the N--connection
coefficients.} There is a class of associated to N--connection local bases, $%
\mathbf{e}_{\nu }=(\mathbf{e}_{i},e_{a}),$ and cobases, $\mathbf{e}^{\mu
}=(e^{i},\mathbf{e}^{a}),$ when
\begin{eqnarray}
\mathbf{e}_{i}&=&\frac{\partial }{\partial x^{i}}-\ N_{i}^{a}(u)\frac{\partial
}{\partial y^{a}}\mbox{ and
}\ e_{a}=\frac{\partial }{\partial y^{a}},  \label{nader} \\
e^{i}&=&dx^{i}\mbox{ and }\mathbf{e}^{a}=dy^{a}+\ N_{i}^{a}(u)dx^{i}.
\label{nadif}
\end{eqnarray}%
Such a structure is, in general, nonholonomic (equivalently, anholonomic/
non--integrable) because, for instance, (\ref{nader}) satisfy nontrivial
nonholonomy relations of type
\begin{equation}
\lbrack \mathbf{e}_{\alpha },\mathbf{e}_{\beta }]=\mathbf{e}_{\alpha }%
\mathbf{e}_{\beta }-\mathbf{e}_{\beta }\mathbf{e}_{\alpha }=W_{\alpha \beta
}^{\gamma }\mathbf{e}_{\gamma },  \label{anhrel}
\end{equation}%
with (antisymmetric) nontrivial anholonomy coefficients $W_{ia}^{b}=\partial
_{a}N_{i}^{b}$ and $W_{ji}^{a}=\Omega _{ij}^{a}$ determined by the
coefficients of curvature of N--connection $\Omega _{ij}^{a}=\mathbf{e}%
_{j}\left( N_{i}^{a}\right) -\mathbf{e}_{i}\left( N_{j}^{a}\right) .$\ It
should be emphasized here that there is a N--connection structure $\mathbf{%
N=\ }^{c}\mathbf{N}$ which is canonically defined by $F.$\footnote{%
Considering $L=F^{2}$ as a regular Lagrangian (i.e. with nondegenerate $%
^{F}g_{ij}$ (\ref{hess})) we can define the action integral $S(\tau
)=\int\limits_{0}^{1}L(x(\tau ),y(\tau ))d\tau $ with $y^{k}(\tau
)=dx^{k}(\tau )/d\tau ,$ for $x(\tau )$ parametrizing smooth curves on $V$ \
with $\tau \in \lbrack 0,1]$. We can prove \cite{ma} that the
Euler--Lagrange equations for $S(\tau ),$ $\frac{d}{d\tau }\frac{\partial L}{%
\partial y^{i}}-\frac{\partial L}{\partial x^{i}}=0,$ are equivalent to the
''nonlinear geodesic'' (equivalently, semi--spray) equations $\frac{%
d^{2}x^{k}}{d\tau ^{2}}+2G^{k}(x,y)=0,$ where $G^{k}=\frac{1}{4}g^{kj}\left(
y^{i}\frac{\partial ^{2}L}{\partial y^{j}\partial x^{i}}-\frac{\partial L}{%
\partial x^{j}}\right) $ induces the canonical N--connection $\mathbf{\ }^{c}%
\mathbf{N=}\{\ ^{c}N_{j}^{a}=\partial G^{a}/\partial y^{j}\}.$} Under
general (co) frame/coordinate transform, $\mathbf{e}^{\alpha }\rightarrow
\mathbf{e}^{\alpha ^{\prime }}=e_{\ \alpha }^{\alpha ^{\prime }}\mathbf{e}%
^{\alpha }$ and/or $u^{\alpha }\rightarrow u^{\alpha ^{\prime }}=u^{\alpha
^{\prime }}(u^{\alpha }),$ preserving the splitting (\ref{whitney}), we get
a corresponding transformation law $\ ^{c}N_{j}^{a}\rightarrow N_{j^{\prime
}}^{a^{\prime }},$ when $\mathbf{N}=N_{i^{\prime }}^{a^{\prime
}}(u)dx^{i^{\prime }}\otimes \frac{\partial }{\partial y^{a^{\prime }}}$ is
given locally by a set of coefficients $\{N_{j}^{a}\}$ (we shall omit
priming, underlying etc of indices if that will not result in ambiguities).%
\footnote{%
We can use any convenient (for constructing exact solutions of field
equations, or geometric considerations) equivalent sets $\mathbf{\ N=}%
\{N_{j}^{a}\},$ which under corresponding frame/coordinate transform can be
parametrized in a form $\ ^{F}\mathbf{N}=\ ^{c}\mathbf{N}=\{\
^{c}N_{j}^{a}\}.$ Here, we also emphasize that we can define conventionally
a N--connection structure on any manifold (not only on tangent/vector
bundles) by prescribing a fibered structure with conventional horizontal (h)
and vertical (v) splitting, for instance, a nonholonomic 2+2 splitting in GR
as we considered in \cite{vex1,vex3}.}

\subsubsection{Finsler metric structure on total tangent bundle}

We can use the so--called Sasaki lift in order to construct on $T\mathbf{V}$
a metric structure completely determined by a fundamental Finsler function $%
F(x,y),$
{\small
\begin{eqnarray}
\ ^{F}\mathbf{g} &=&(h\ ^{F}g_{ij},v\ ^{F}g_{ij})
=\ ^{F}g_{ij}(x,y)[\ e^{i}\otimes e^{j}+\left( \ ^{\ast }\mathit{l}%
_{P}\right) ^{2}\ ^{F}\mathbf{e}^{i}\otimes \ \ ^{F}%
\mathbf{e}^{j}],  \label{slm} \\
e^{i} &=&dx^{i}\mbox{ and }\ ^{F}\mathbf{e}^{a}=dy^{a}+\
^{F}N_{i}^{a}(u)dx^{i},  \label{ddifl}
\end{eqnarray}
}%
where for canonical constructions $\ ^{F}N=\mathbf{\ }^{c}\mathbf{N.}$ In
the above formula we consider a length constant $\ ^{\ast }\mathit{l}_{P}$
which can be just the Planck length $\mathit{l}_{P}$ in models of QG for the
GR but it can be a different one for brane models. We have to consider such
a value before the v--part of metric (\ref{slm}) in order to have the same
dimensions for the h-- and v--components of metric when coordinates have the
dimensions $[x^{i}]=cm$ and $[y^{i}\sim dx^{i}/ds]=cm/cm.$ In our further
considerations, we shall include such a constant into $h$--coefficients of
metrics if that will not result in ambiguities.

Under general frame transforms $e^{\alpha ^{\prime }}=e_{\ \alpha }^{\alpha
^{\prime }}\mathbf{e}^{\alpha },$ the above Finsler metric can be
represented in a general 4+4 form
\begin{eqnarray}
\ \ ^{H}\mathbf{g} &=&(h\ g_{ij},vg_{ab})=\ \ ^{H}g_{\alpha \beta }(x,y)\
\mathbf{e}^{\alpha }\otimes \ \mathbf{e}^{\beta }  \label{dm} \\
&=&\ g_{ij}(x,y)\ e^{i}\otimes e^{j}+\left( \ ^{\ast }\mathit{l}_{P}\right)
^{2}\ h_{ab}(x,y)\ \mathbf{e}^{a}\otimes \ \mathbf{e}^{b},  \notag
\end{eqnarray}%
for arbitrary $N_{j^{\prime }}^{a}$ (we put the left label $H$ in order to
emphasize that such a metric is induces by MDR and nonholonomic deformations
from HL gravity). With respect to a coordinate co-basis $du^{\beta
}=(dx^{j},dy^{b}),$ when $\partial _{\alpha }=\partial /\partial u^{\alpha
}=(\partial _{i}=\partial /\partial x^{i},\partial _{a}=\partial /\partial
y^{a}),$ both metrics can be transformed equivalently into
\begin{equation}
\ \ ^{H}\mathbf{g}=\ \ ^{H}\ \underline{g}_{\alpha \beta }\left( u\right)
du^{\alpha }\otimes du^{\beta },  \label{fmetr}
\end{equation}%
where%
\begin{equation}
\ \ ^{H}\ \underline{g}_{\alpha \beta }=\left[
\begin{array}{cc}
\ g_{ij}+\left( \ ^{\ast }\mathit{l}_{P}\right) ^{2}\
h_{ab}N_{i}^{a}N_{j}^{b} & \left( \ ^{\ast }\mathit{l}_{P}\right) ^{2}\
h_{ae}N_{j}^{e} \\
\ \left( \ ^{\ast }\mathit{l}_{P}\right) ^{2}\ h_{be}N_{i}^{e} & \left( \
^{\ast }\mathit{l}_{P}\right) ^{2}\ \ h_{ab}%
\end{array}%
\right] .  \label{fansatz}
\end{equation}%
The values $N_{i}^{a}(u)$ should be not identified to certain gauge fields
in a Kaluza--Klein theory on tangent bundle with the potentials depending on
velocities if we do not consider compactifications on coordinates $y^{a}.$
In Finsler like theories, a set $\{N_{i}^{a}\}$ defines a N--connection
structure, with elongated partial derivatives (\ref{ddifl}).

We can invert the constructions for arbitrary (\ref{fmetr}) and/or (\ref{dm}%
) and introduce Finsler variables and define metric (\ref{slm}) by
prescribing an arbitrary generating function $F$ on a manifold or bundle
space.

\subsubsection{The canonical distinguished Finsler connection}

In order to perform self--consistent geometric constructions with h-- and
v--splitting, it was introduced the concept of distinguished connection (in
brief, d--connection). A d--connection $\mathbf{D=(}h\mathbf{D,}v\mathbf{D)}$
is defined as a linear one \ preserving under parallelism the N--connection
structure on $\mathbf{V.}$ The N--adapted components $\mathbf{\Gamma }_{\
\beta \gamma }^{\alpha }$ of a d--connection $\mathbf{D}$ are computed
following equations $\mathbf{D}_{\alpha }\mathbf{e}_{\beta }=\mathbf{\Gamma }%
_{\ \alpha \beta }^{\gamma }\mathbf{e}_{\gamma }$ and parametrized in the
form $\ \mathbf{\Gamma }_{\ \alpha \beta }^{\gamma }=\left(
L_{jk}^{i},L_{bk}^{a},C_{jc}^{i},C_{bc}^{a}\right) ,$ where $\mathbf{D}%
_{\alpha }=(D_{i},D_{a}),$ with $h\mathbf{D}=(L_{jk}^{i},L_{bk}^{a})$ and $v%
\mathbf{D}=(C_{jc}^{i},$ $C_{bc}^{a})$ defining certain covariant,
respectively, h-- and v--derivatives.

The simplest way to perform computations with a d--connection $\mathbf{D}$
is to associate it with a N--adapted differential 1--form
\begin{equation}
\mathbf{\Gamma }_{\ \beta }^{\alpha }=\mathbf{\Gamma }_{\ \beta \gamma
}^{\alpha }\mathbf{e}^{\gamma },  \label{dconf}
\end{equation}%
and apply on $T\mathbf{V}$ the well known formalism of differential forms as
in GR. For instance, the torsion of $\ \mathbf{D}$ is defined/computed
\begin{equation}
\mathcal{T}^{\alpha }\doteqdot \mathbf{De}^{\alpha }=d\mathbf{e}^{\alpha }+%
\mathbf{\Gamma }_{\ \beta }^{\alpha }\wedge \mathbf{e}^{\beta }.
\label{tors}
\end{equation}%
With respect to a N--adapted basis, this torsion is stated by $\mathcal{T}=\{%
\mathbf{T}_{\ \alpha \beta }^{\gamma }\equiv \mathbf{\Gamma }_{\ \alpha
\beta }^{\gamma }-\mathbf{\Gamma }_{\ \beta \alpha }^{\gamma };T_{\
jk}^{i},T_{\ ja}^{i},T_{\ ji}^{a},T_{\ bi}^{a},T_{\ bc}^{a}\},$ where the
nontrivial coefficients are
\begin{eqnarray}
T_{\ jk}^{i} &=&L_{jk}^{i}-L_{kj}^{i},T_{\ ja}^{i}=C_{jb}^{i},T_{\
ji}^{a}=-\Omega _{\ ji}^{a},  \label{dtors} \\
T_{aj}^{c} &=&L_{aj}^{c}-e_{a}(N_{j}^{c}),T_{\ bc}^{a}=C_{bc}^{a}-C_{cb}^{a}.
\notag
\end{eqnarray}

There is a canonical d--connection\footnote{%
For any of type of metric parametrizatons (\ref{dm}), (\ref{fmetr}) and/or (%
\ref{slm}), we can construct the Levi--Civita connection $\nabla =\{\Gamma
_{\ \beta \gamma }^{\alpha }\}$ on $\mathbf{V}$ in a standard form. This
connection is not used in Finsler geometry and generalizations because it is
not compatible with a N--connection splitting; under parallel transports
with $\nabla ,$ it is not preserved the Whitney sum (\ref{whitney}).}, $\
^{H}$ $\widehat{\mathbf{D}}=\{\ ^{H}\widehat{\mathbf{\Gamma }}_{\ \alpha
\beta }^{\gamma }=(\widehat{L}_{jk}^{i},\widehat{L}_{bk}^{a},\widehat{C}%
_{jc}^{i},$ $\widehat{C}_{bc}^{a})\},$ which is uniquely and completely
defined by the coefficients of metric $\mathbf{g}$ (\ref{dm}) (equivalently,
(\ref{fmetr}) and/or (\ref{slm})) following the metric compatibility
conditions that $\widehat{\mathbf{D}}\mathbf{g}=0$ and the ''pure''
horizontal and vertical torsion coefficients are zero, i. e. $\widehat{T}_{\
jk}^{i}=0$ and $\widehat{T}_{\ bc}^{a}=0,$
\begin{eqnarray}
\widehat{L}_{jk}^{i} &=&\frac{1}{2}g^{ir}\left(
e_{k}g_{jr}+e_{j}g_{kr}-e_{r}g_{jk}\right) ,  \label{candcon} \\
\widehat{L}_{bk}^{a} &=&e_{b}(N_{k}^{a})+\frac{1}{2}h^{ac}\left(
e_{k}h_{bc}-h_{dc}\ e_{b}N_{k}^{d}-h_{db}\ e_{c}N_{k}^{d}\right) ,  \notag \\
\widehat{C}_{jc}^{i} &=&\frac{1}{2}g^{ik}e_{c}g_{jk},\ \widehat{C}_{bc}^{a}=%
\frac{1}{2}h^{ad}\left( e_{c}h_{bd}+e_{c}h_{cd}-e_{d}h_{bc}\right) .  \notag
\end{eqnarray}%
Such a d--connection contains nontrivial torsion components $\widehat{T}_{\
ja}^{i},\widehat{T}_{\ ji}^{a},\widehat{T}_{aj}^{c},$ i.e., in general, $\ \
^{H}\widehat{\mathcal{T}}\neq 0.$ It is very different from various types of
torsions in Einstein--Cartan, gauge, string and other type gravity theories
(for which additional field equations are defined) because its N--adapted
components are completely by the metric structure, which in its turn (in our
model) is related to MDR in HL\ gravity - we do not need additional field
equations for this type of torsions induced nonholonomically via the
N--connection structure.\footnote{%
Any geometric/physical construction for $\widehat{\mathbf{D}}$ can be
re--defined equivalently into a similar one with the Levi--Civita connection
following formula
\begin{equation*}
\Gamma _{\ \alpha \beta }^{\gamma }=\widehat{\mathbf{\Gamma }}_{\ \alpha
\beta }^{\gamma }+Z_{\ \alpha \beta }^{\gamma },
\end{equation*}%
where the distortion tensor $Z_{\ \alpha \beta }^{\gamma }$ is given by
nontrivial coefficients%
\begin{eqnarray*}
\ Z_{jk}^{a} &=&-\widehat{C}_{jb}^{i}g_{ik}h^{ab}-\frac{1}{2}\Omega
_{jk}^{a},~Z_{bk}^{i}=\frac{1}{2}\Omega _{jk}^{c}h_{cb}g^{ji}-\Xi _{jk}^{ih}~%
\widehat{C}_{hb}^{j}, \\
Z_{bk}^{a} &=&~^{+}\Xi _{cd}^{ab}~\widehat{T}_{kb}^{c},\ Z_{kb}^{i}=\frac{1}{%
2}\Omega _{jk}^{a}h_{cb}g^{ji}+\Xi _{jk}^{ih}~\widehat{C}_{hb}^{j},\
Z_{jk}^{i}=0, \\
\ Z_{jb}^{a} &=&-~^{-}\Xi _{cb}^{ad}~\widehat{T}_{jd}^{c},\ Z_{bc}^{a}=0,\
Z_{ab}^{i}=-\frac{g^{ij}}{2}\left[ \widehat{T}_{ja}^{c}h_{cb}+\widehat{T}%
_{jb}^{c}h_{ca}\right] ,
\end{eqnarray*}%
for $\ \Xi _{jk}^{ih}=\frac{1}{2}(\delta _{j}^{i}\delta
_{k}^{h}-g_{jk}g^{ih})$ and $~^{\pm }\Xi _{cd}^{ab}=\frac{1}{2}(\delta
_{c}^{a}\delta _{d}^{b}+h_{cd}h^{ab}).$}

Via nonholonomic transforms, we can transform $\ ^{H}\widehat{\mathbf{D}}$
into the Cartan d--con\-nection $\ ^{H}\widetilde{\mathbf{D}}$ in Finsler
geometry which is also metric compatible and completely defined by the same
metric structure. On spaces of even dimensions such connections contain the
same physical information if a Finsler generating function $F$ on spacetime
manifold. The metric compatibility play a crucial role in defining Finsler
generalizations of gravity in an "almost standard form" following principles
which are similar to those in GR (it is a more sophisticate task to
elaborate viable physical models using metric noncompatible connections, for
instance, the so--called Chern connection for Finsler geometry, see critical
remarks and details in Refs. \cite{vcrit,vcosm,vrev1}).

\subsubsection{Nonholonomic deformations relating HF and HL metrics}

The \textbf{Horava--Finsler }(HF) gravity theory is a (pseudo) Finsler
geometry model induced canonically on $T\mathbf{V}$ by a Finsler generating
function $F$ associated to MDE relations in ''standard''\footnote{%
the word standard is an approximation because up till present there are
different versions of HL with, or not, detailed balance conditions,
generalized forms etc} HL gravity. Such a theory is determined by the data $%
[F:\ ^{F}\mathbf{g=}(h\ ^{F}g_{ij},v\ ^{F}g_{ij})\mathbf{,\ }^{F}\mathbf{N,\
}^{F}\mathbf{D=}\ \ ^{H}\widehat{\mathbf{D}}],$ where (up to frame
transforms)
\begin{equation*}
\ ^{F}g_{ij}(x,y)\sim g_{ij}(x,y)=e_{\ i}^{i^{\prime }}(x,y)e_{\
j}^{j^{\prime }}(x,y)\ ^{HL}g_{ij}(x),
\end{equation*}
for $\ ^{HL}g_{ij}(x)$ being a solution of gravitational field equations in
HL\ gravity on $\mathbf{V,}$ derived from action (\ref{action}). The values $%
e_{\ i}^{i^{\prime }}(x,y)$ and $h_{bc}(x,y)$ have to be defined from
certain solutions of gravitational field equations in HF gravity, see next
section \ref{ssfeq}.

If the conditions (\ref{grcond}) are imposed in HF gravity, we can state
such limits that the model defines an \textbf{Einstein--Finsler gravity}
theory (EFG) \cite{vcosm}. This class of metric compatible Finsler gravity
theories on $TV$ is defined by data $[F:\ ^{F}\mathbf{g=}(h\ ^{F}g_{ij},v\
^{F}g_{ij})\mathbf{,\ }^{F}\mathbf{N,\ }^{F}\mathbf{D=}\widehat{\mathbf{D}}%
], $ when $\ ^{F}g_{ij}(x,y)\sim g_{ij}(x,y)=e_{\ i}^{i^{\prime }}(x,y)e_{\
j}^{j^{\prime }}(x,y)\ ^{E}g_{ij}(x),$ for $\ ^{E}g_{ij}(x)$ being a
solution of the Einstein equations in GR. Via nonholonomic frame transforms,
the theory can be equivalently described in standard variables of GR with $%
\left[ \mathbf{g},\ ^{g}\nabla \right].$

In explicit form, we have to elaborate a natural trapping/warped mechanisms
defined by explicit solutions of (Finsler type) gravitational field
equations which in classical limits for $\ ^{\ast }\mathit{l}_{P}\rightarrow
0,$ when HF / EFG $\rightarrow $ HL, or GR, determining QG corrections to
gravitational and matter field interactions at different scales depending on
the class of considered models and solutions, see below section \ref{sec4}.

\subsection{Field equations in canonical Ho\v{r}ava--Finsler gravity}

\label{ssfeq}A canonical (pseudo) Finsler structure on $T\mathbf{V}$
determined by MDR in HL gravity contains already all anisotropic properties
which are contained in metrics (\ref{adm}) with scaling properties (\ref%
{scalin}) included in the $h$--part of corresponding N--adapted metric (\ref%
{dm}) and/or (\ref{slm}). We elaborate a Ho\v rava--Finsler gravity theory
not just lifting formally the geometric objects and action (\ref{action}) on
$T\mathbf{V}$ (geometrically such a procedure can be defined in a canonical
way). There are not experimental data about matter fields and their
energy--momentums on tangent/vector bundles. A ''simple'' approach is to
develop a Finsler brane gravity model with general assumptions on matter
field in the bulk and warping/trapping of matter on a 4--d base spacetime
(see details in Refs. \cite{vbrane} and, for some yearly off--diagonal
constructions with N--connection structure for generalized Rundall--Sundrum
scenarios, \cite{vsingl1,vsingl2}, and references therein). A well--defined
trapping mechanism with effective (in general, anisotropically polarized)
cosmological constant and maximal speed of light (as solutions for the bulk
HF gravity) allows us to simplify substantially the constructions related to
possible models of HF gravity which can transform in the quasi--classical
limit into the HL or/and GR theories.

Using the canonical d--connection 1--form of type (\ref{dconf}), with
coefficients $\ \ ^{H}\widehat{\mathbf{\Gamma }}_{\ \alpha \beta }^{\gamma }$
(\ref{candcon}), we can compute the curvature of $\ \ \ ^{H}\widehat{\mathbf{%
D}},$
\begin{equation}
\ \ ^{H}\widehat{\mathcal{R}}_{~\beta }^{\alpha }:=\ \ ^{H}\widehat{\mathbf{D%
}}\ \ ^{H}\widehat{\mathbf{\Gamma }}_{\ \beta }^{\alpha }=d\ \ ^{H}\widehat{%
\mathbf{\Gamma }}_{\ \beta }^{\alpha }-\ \ ^{H}\widehat{\mathbf{\Gamma }}_{\
\beta }^{\gamma }\wedge \ \ ^{H}\widehat{\mathbf{\Gamma }}_{\ \gamma
}^{\alpha }=\widehat{\mathbf{R}}_{\ \beta \gamma \delta }^{\alpha }\mathbf{e}%
^{\gamma }\wedge \mathbf{e}^{\delta },  \label{curv}
\end{equation}%
see details in \cite{vcosm,vrev1}, where the formulas for all coefficients
are given in explicit form. The Ricci d--tensor $\widehat{R}ic=\{\widehat{%
\mathbf{R}}_{\alpha \beta }\}$ is defined by contracting respectively the
components of curvature tensor, $\widehat{\mathbf{R}}_{\alpha \beta
}\doteqdot \widehat{\mathbf{R}}_{\ \alpha \beta \tau }^{\tau },$ The h--/
v--components of this d--tensor, $\widehat{\mathbf{R}}_{\alpha \beta }=\{%
\widehat{R}_{ij},\widehat{R}_{ia},\ \widehat{R}_{ai},\ \widehat{R}_{ab}\},$
are
\begin{equation}
\widehat{R}_{ij}:=\widehat{R}_{\ ijk}^{k},\ \ \widehat{R}_{ia}:=-\widehat{R}%
_{\ ika}^{k},\ \widehat{R}_{ai}:=\widehat{R}_{\ aib}^{b},\ \widehat{R}_{ab}:=%
\widehat{R}_{\ abc}^{c}.  \label{dricci}
\end{equation}

The scalar curvature of $\ \ ^{H}\widehat{\mathbf{D}}$ is constructed by
using the inverse to $\mathbf{g}$ (\ref{dm}),
\begin{equation}
\ ^{s}\widehat{\mathbf{R}}:=\mathbf{g}^{\alpha \beta }\widehat{\mathbf{R}}%
_{\alpha \beta }=g^{ij}\widehat{R}_{ij}+h^{ab}\widehat{R}_{ab}=\check{R}+%
\check{S},  \label{sdccurv}
\end{equation}%
where $\check{R}=g^{ij}\widehat{R}_{ij}$ and $\check{S}=h^{ab}\widehat{R}%
_{ab}$ are respectively the h-- and v--components of scalar curvature.

The Einstein tensor for$\ \ ^{H}\widehat{\mathbf{D}}$ \ is, by definition,
\begin{equation}
\ \ ^{H}\widehat{\mathbf{E}}_{\alpha \beta }:=\widehat{\mathbf{R}}_{\alpha
\beta }-\frac{1}{2}\mathbf{g}_{\alpha \beta }\ ^{s}\widehat{\mathbf{R}}.
\label{enstdt}
\end{equation}%
We can postulate the gravitational field equation for the HF gravity on $T%
\mathbf{V}$ in the form
\begin{equation}
\ \ ^{H}\widehat{\mathbf{E}}_{\alpha \beta }=\widehat{\mathbf{\Upsilon }}%
_{\beta \delta },  \label{ensteqcdc}
\end{equation}%
for arbitrary sources $\widehat{\mathbf{\Upsilon }}_{\beta \delta }$ which
can be, as a matter of principle, defined as certain lifts of
energy--momentum tensors of matter fields in HL, or GR, theory. It should be
emphasized here that the solutions of equations (\ref{ensteqcdc}), for
''projections'' $T\mathbf{V\rightarrow V,}$ in general, do not transform
trivially into solutions of HL gravity with action (\ref{action}). Certain
warped/trapping scenarios can be constructed in such a form that
nonholonomic deformations of exact solution in HF brane gravity are, in
general, non--explicitly related to solutions in HL gravity. This is a
consequence of nontrivial nonholonomic structure and generic nonlinear
character of such locally anisotropic gravitational systems.

\subsection{Magic splitting of gravitational HF filed equations}

The gravitational field equations in HF gravity can be integrated in very
general forms on $T\mathbf{V}$ following the anholonomic deformation method
summarized in Refs. \cite{vex1,vex2,vex3} (necessary ''velocity'' type
coordinated should be treated as certain ''extra'' dimension to two/four
dimensional base space ones). The bulk of such solutions do not have obvious
implications in modern physics. For simplicity, in this work we shall use a
more restricted class of exact solutions in HF gravity which seem to be
related to models of Finsler branes.

We parametrize the metric (\ref{dm}) in a form with three ''shell''
anisotro\-py (with a nonholonomic splitting 2+2 and 2+2+2),
\begin{eqnarray}
\ \ \mathbf{g} &=&\ g_{ij}(x)dx^{i}\otimes dx^{j}+h_{\ ^{0}a\ ^{0}b}(x,\
^{0}y)\mathbf{e}^{\ ^{0}a}\otimes \mathbf{e}^{\ ^{0}b}  \label{2forman} \\
&&+h_{\ ^{1}a_{1}\ ^{1}b}(x,\ ^{0}y,\ ^{1}y)\mathbf{e}^{\ ^{1}a}\otimes
\mathbf{e}^{\ ^{1}b}+h_{\ ^{2}a\ ^{2}b}(x,\ ^{0}y,\ ^{1}y,\ ^{2}y)\mathbf{e}%
^{\ ^{2}a}\otimes \mathbf{e}^{\ ^{2}b},  \notag \\
\mathbf{e}^{\ ^{0}a} &=&dy^{\ ^{0}a}+N_{i}^{\ ^{0}a}(\ ^{0}u)dx^{i},  \notag
\\
\mathbf{e}^{\ ^{1}a} &=&dy^{\ ^{1}a}+N_{i}^{\ ^{1}a}(\ ^{1}u)dx^{i}+N_{\
^{0}a}^{\ ^{1}a}(\ ^{1}u)\ \mathbf{e}^{\ ^{0}a},  \notag \\
\mathbf{e}^{\ ^{2}a} &=&dy^{\ ^{2}a}+N_{i}^{\ ^{2}a}(\ ^{2}u)dx^{i}+N_{\
^{0}a}^{\ ^{0}a}(\ ^{2}u)\ \mathbf{e}^{\ ^{0}a}+N_{\ ^{1}a}^{\ ^{2}a}(\
^{2}u)\ \mathbf{e}^{\ ^{1}a},  \notag
\end{eqnarray}%
for local $x=\{x^{i}\},\ ^{0}y=\{y^{\ ^{0}a}\},\ ^{1}y=\{y^{\ ^{1}a}\},\
^{2}y=\{y^{\ ^{2}a}\};$ the vertical indices and coordinates split in the
form $y=[\ ^{0}y,\ ^{1}y,\ ^{2}y],$ or $y^{a}=[y^{\ ^{0}a},y^{\ ^{1}a},y^{\
^{2}a}];$ $\ ^{0}u=(x,\ ^{0}y),\ ^{1}u=(\ ^{0}u,\ ^{1}y),\ ^{2}u=(\ ^{1}u,\
^{2}y),$ or $u^{\ \alpha }=u^{\ ^{0}\alpha }=(x^{i},y^{\ ^{0}a}),u^{\
^{1}\alpha }=(u^{\ ^{0}\alpha },y^{\ ^{1}a}),u^{\ ^{2}\alpha }=(u^{\
^{1}\alpha },y^{\ ^{2}a}).$ There is a ''less'' general ansatz of type (\ref%
{2forman}) (with Killing symmetry on $y^{8},$ when the metric coefficients
do not depend on variable $y^{8};$ it is convenient to write $\ y^{3}=\
^{0}v,$ $y^{5}=\ ^{1}v,y^{7}=\ ^{2}v$ and express the $N$--coefficients via $%
n$-- and $w$--functions)
\begin{eqnarray}
\ ^{sol}\mathbf{g} &=&g_{i}(x^{k})dx^{i}\otimes dx^{j}+h_{\ ^{0}a}(x^{k},\
^{0}v)\mathbf{e}^{\ ^{0}a}{\otimes }\mathbf{e}^{\ ^{0}a}  \label{ansgensol}
\\
&&+h_{\ ^{1}a}(u^{\ ^{0}\alpha },\ ^{1}v)\ \mathbf{e}^{\ ^{1}a}{\otimes }\
\mathbf{e}^{\ ^{1}a}+h_{\ ^{2}a}(u^{\ ^{1}\alpha },\ ^{2}v)\ \mathbf{e}^{\
^{2}a}{\otimes }\ \mathbf{e}^{\ ^{2}a},  \notag
\end{eqnarray}%
\begin{eqnarray}
\mathbf{e}^{3} &=&dy^{3}+w_{i}(x^{k},\ ^{0}v)dx^{i},\mathbf{e}%
^{4}=dy^{4}+n_{i}(x^{k},\ ^{0}v)dx^{i},  \notag \\
\mathbf{e}^{5} &=&dy^{5}+w_{\ ^{0}\beta }(u^{\ ^{0}\alpha },\
^{1}v)du^{\beta },\mathbf{e}^{6}=dy^{6}+n_{\ ^{0}\beta }(u^{\ ^{0}\alpha },\
^{1}v)du^{\ ^{0}\beta },  \notag \\
\mathbf{e}^{7} &=&dy^{7}+w_{\ ^{1}\beta }(u^{\ ^{1}\alpha },\ ^{2}v)du^{\
^{1}\beta },\mathbf{e}^{8}=dy^{8}+n_{\ ^{1}\beta }(u^{\ ^{1}\alpha },\
^{2}v)du^{\ ^{1}\beta }.  \notag
\end{eqnarray}

The HF gravitational field equations (\ref{ensteqcdc}) for the canonical
d--connection $\widehat{\mathbf{D}}$ can be solved in general forms for
ansatz (\ref{ansgensol}) and sources parametrized with respect N--adapted
frames in the form
\begin{eqnarray}
\widehat{\mathbf{\Upsilon }}_{\ \delta }^{\beta } &=&diag[\widehat{\mathbf{%
\Upsilon }}_{\ 1}^{1}=\widehat{\mathbf{\Upsilon }}_{\ 2}^{2}=\widehat{%
\mathbf{\Upsilon }}_{2}(u^{\ ^{2}\alpha }),\widehat{\mathbf{\Upsilon }}_{\
3}^{3}=\widehat{\mathbf{\Upsilon }}_{\ 4}^{4}=\widehat{\mathbf{\Upsilon }}%
_{4}(u^{\ ^{2}\alpha }),  \notag \\
&&\widehat{\mathbf{\Upsilon }}_{\ 5}^{5}=\widehat{\mathbf{\Upsilon }}_{\
6}^{6}=\widehat{\mathbf{\Upsilon }}_{6}(u^{\ ^{2}\alpha }),\widehat{\mathbf{%
\Upsilon }}_{\ 7}^{7}=\widehat{\mathbf{\Upsilon }}_{\ 8}^{8}=\widehat{%
\mathbf{\Upsilon }}_{8}(u^{\ ^{2}\alpha })],  \label{sourcb}
\end{eqnarray}%
when the coefficients are subjected to algebraic conditions (for vanishing
N---coefficients, containing respectively the functions (\ref{source3})
determining sources in the gravitational field equations)%
\begin{eqnarray}
\ ^{h}\Lambda (x^{i}) &=&\widehat{\mathbf{\Upsilon }}_{4}+\widehat{\mathbf{%
\Upsilon }}_{6}+\widehat{\mathbf{\Upsilon }}_{8},\ ^{v}\Lambda (x^{i},v)=%
\widehat{\mathbf{\Upsilon }}_{2}+\widehat{\mathbf{\Upsilon }}_{6}+\widehat{%
\mathbf{\Upsilon }}_{8},  \label{sourcb1} \\
\ ^{1}\Lambda \ (u^{\ \alpha },y^{5}) &=&\widehat{\mathbf{\Upsilon }}_{2}+%
\widehat{\mathbf{\Upsilon }}_{4}+\widehat{\mathbf{\Upsilon }}_{8},\ \ \
^{2}\Lambda (u^{\ \ ^{1}\alpha },y^{7})\ =\widehat{\mathbf{\Upsilon }}_{2}+%
\widehat{\mathbf{\Upsilon }}_{4}+\widehat{\mathbf{\Upsilon }}_{6}.\   \notag
\end{eqnarray}%
Introducing the coefficients of metric (\ref{ansgensol}) into the formulas
for d--connection (\ref{candcon}) after tedious calculations (see details in %
\cite{vex1,vex2})) we obtain \

\begin{eqnarray}
\widehat{R}_{1}^{1} &=&\widehat{R}_{2}^{2}  \label{eqe1} \\
&=&-\frac{1}{2g_{1}g_{2}}\left[ g_{2}^{\bullet \bullet }-\frac{%
g_{1}^{\bullet }g_{2}^{\bullet }}{2g_{1}}-\frac{\left( g_{2}^{\bullet
}\right) ^{2}}{2g_{2}}+g_{1}^{\prime \prime }-\frac{g_{1}^{\prime
}g_{2}^{\prime }}{2g_{2}}-\frac{\left( g_{1}^{\prime }\right) ^{2}}{2g_{1}}%
\right] =-\ ^{h}\Lambda (x^{k}),  \notag \\
\widehat{R}_{3}^{3} &=&\widehat{R}_{4}^{4}=-\frac{1}{2h_{3}h_{4}}\left[
h_{4}^{\ast \ast }-\frac{\left( h_{4}^{\ast }\right) ^{2}}{2h_{4}}-\frac{%
h_{3}^{\ast }h_{4}^{\ast }}{2h_{3}}\right] =-\ ^{v}\Lambda (x^{k},y^{3}),
\label{eqe2}
\end{eqnarray}%
\begin{eqnarray}
\widehat{R}_{3k}=\frac{w_{k}}{2h_{4}}[h_{4}^{\ast \ast }-\frac{(h_{4}^{\ast
})^{2}}{2h_{4}}-\frac{h_{3}^{\ast }h_{4}^{\ast }}{2h_{3}}]+\frac{h_{4}^{\ast
}}{4h_{4}}(\frac{\partial _{k}h_{3}}{h_{3}}+\frac{\partial _{k}h_{4}}{h_{4}}%
)-\frac{\partial _{k}h_{4}^{\ast }}{2h_{4}}=0 &&  \label{eqe3a} \\
\widehat{R}_{4k}=\frac{h_{4}}{2h_{3}}n_{k}^{\ast \ast }+\left( \frac{h_{4}}{%
h_{3}}h_{3}^{\ast }-\frac{3}{2}h_{4}^{\ast }\right) \frac{n_{k}^{\ast }}{%
2h_{3}}=0, &&  \label{eqe3}
\end{eqnarray}%
where certain differential derivatives are denoted in the form $a^{\bullet
}=\partial a/\partial x^{1},$\ $a^{\prime }=\partial a/\partial x^{2},$\ $%
a^{\ast }=\partial a/\partial y^{3},$ and (extra to 4--d ''shell''
equations)
\begin{eqnarray*}
\widehat{R}_{5}^{5} &=&\widehat{R}_{6}^{6}=-\frac{1}{2h_{5}h_{6}}\left[
\partial _{\ ^{1}v\ ^{1}v}^{2}h_{6}-\frac{\left( \partial _{\
^{1}v}h_{6}\right) ^{2}}{2h_{6}}-\frac{(\partial _{\ ^{1}v}h_{5})(\partial
_{\ ^{1}v}h_{6})}{2h_{5}}\right] \\
&&=-\ \ ^{1}\Lambda \ (u^{\ \alpha },y^{5}), \\
\widehat{R}_{7}^{7} &=&\widehat{R}_{8}^{8}=-\frac{1}{2h_{7}h_{8}}\left[
\partial _{\ ^{2}v\ ^{2}v}^{2}h_{8}-\frac{\left( \partial _{\
^{2}v}h_{8}\right) ^{2}}{2h_{8}}-\frac{(\partial _{\ ^{2}v}h_{7})(\partial
_{\ ^{2}v}h_{6})}{2h_{7}}\right] \\
&& =-\ ^{2}\Lambda (u^{\ \ ^{1}\alpha },y^{7}),
\end{eqnarray*}%
\begin{eqnarray}
\widehat{R}_{5\ ^{0}\alpha } &=&\frac{\ ^{1}w_{\ ^{0}\alpha }}{2h_{6}}\left[
\partial _{\ ^{1}v\ ^{1}v}^{2}h_{6}-\frac{\left( \partial _{\
^{1}v}h_{6}\right) ^{2}}{2h_{6}}-\frac{(\partial _{\ ^{1}v}h_{5})(\partial
_{\ ^{1}v}h_{6})}{2h_{5}}\right]  \notag \\
&&+\frac{\partial _{\ ^{1}v}h_{6}}{4h_{6}}\left( \frac{\partial _{\
^{0}\alpha }h_{5}}{h_{5}}+\frac{\partial _{\ ^{0}\alpha }h_{6}}{h_{6}}%
\right) -\frac{\partial _{\ ^{0}\alpha }\partial _{\ ^{1}v}h_{6}}{2h_{6}}=0,
\label{eqeho} \\
\widehat{R}_{6\ ^{0}\alpha } &=&\frac{h_{6}}{2h_{5}}\partial _{\ ^{1}v\
^{1}v}^{2}\ ^{1}n_{\ ^{0}\alpha }+\left( \frac{h_{6}}{h_{5}}\partial _{\
^{1}v}h_{5}-\frac{3}{2}\partial _{\ ^{1}v}h_{6}\right) \frac{\partial _{\
^{1}v}\ ^{1}n_{\ ^{0}\alpha }}{2h_{5}}=0,  \notag
\end{eqnarray}%
\begin{eqnarray*}
\widehat{R}_{7\ ^{1}\alpha } &=&\frac{\ ^{2}w_{\ ^{1}\alpha }}{2h_{4}}\left[
\partial _{\ ^{2}v\ ^{2}v}^{2}h_{8}-\frac{\left( \partial _{\
^{2}v}h_{8}\right) ^{2}}{2h_{8}}-\frac{(\partial _{\ ^{2}v}h_{7})(\partial
_{\ ^{2}v}h_{8})}{2h_{7}}\right] \\
&&+\frac{(\partial _{\ ^{2}v}h_{8})}{4h_{8}}\left( \frac{\partial _{\
^{1}\alpha }h_{7}}{h_{7}}+\frac{\partial _{\ ^{1}\alpha }h_{8}}{h_{8}}%
\right) -\frac{\partial _{\ ^{1}\alpha }\partial _{\ ^{2}v}h_{8}}{2h_{8}}=0,
\\
\widehat{R}_{8\ ^{1}\alpha } &=&\frac{h_{8}}{2h_{7}}\partial _{\ ^{2}v\
^{2}v}^{2}\ ^{2}n_{\ ^{1}\alpha }+\left( \frac{h_{8}}{h_{7}}\partial _{\
^{2}v}h_{7}-\frac{3}{2}\partial _{\ ^{2}v}h_{8}\right) \frac{\partial _{\
^{2}v}\ ^{2}n_{\ ^{1}\alpha }}{2h_{8}}=0,
\end{eqnarray*}%
where partial derivatives, for instance, are $\partial _{\ ^{1}v}=\partial
/\partial \ ^{1}v=$ $\partial /\partial y^{5},\ $ $\partial _{\
^{2}v}=\partial /\partial \ ^{2}v=$ $\partial /\partial y^{7},$ and $N_{\
^{0}\alpha }^{5}=\ ^{1}w_{\ ^{0}\alpha }(u^{\ ^{0}\alpha },\ ^{1}v),\ N_{\
^{0}\alpha }^{6}=\ ^{1}n_{\ ^{0}\alpha }(u^{\ ^{0}\alpha },\ ^{1}v),$ $N_{\
^{1}\alpha }^{7}=\ ^{2}w_{\ ^{1}\alpha }(u^{\ ^{1}\alpha },\ ^{2}v),\ N_{\
^{1}\alpha }^{8}=\ ^{2}n_{\ ^{1}\alpha }(u^{\ ^{1}\alpha },\ ^{2}v).$

The above system of equations is a generic nonlinear one with partial
derivatives. Surprisingly, the existing separation of equations (we should
not confuse with separation of variables which is a different property)
allows us to construct very general classes of exact solutions (depending on
the conditions if certain partial derivatives are zero, or not), see
detailed analysis, discussions possible applications in modern gravity and
cosmology in \cite{vex2,vcosm,vrev1}.

Let us explain using the set of equations (\ref{eqe1})--(\ref{eqe3}) the
property of separation of equations for ansatz of type (\ref{ansgensol}).
For a HL model with given matter fields on $\mathbf{V,}$ we construct the
energy momentum tensor $T_{ij}$.\ We can consider a nonholonomic lift on $T%
\mathbf{V}$ such way organized that the resulting in $\Upsilon
_{j}^{i}=diag[Y_{1}^{1}=Y_{2}^{2}=\Upsilon
_{4}(x^{k},y^{3}),Y_{3}^{3}=Y_{4}^{4}=\Upsilon _{2}(x^{k})]$ (using
corresponding noholonomic distributions and transforms, various types of
physically motivated energy--momentum tensors can be parametrized in such a
diagonal form with respect to N--adapted frames)$.$ Taking the value $%
\Upsilon _{2}(x^{k}),$ we can define $g_{1}(x^{k})$ (or, inversely, $%
g_{2}(x^{k})$) for a given $g_{2}(x^{k})$ (or, inversely, $g_{1}(x^{k})$) as
an explicit, or non--explicit, solution of (\ref{eqe1}) by integrating two
times on $h$--variables. Similarly, for a given $\Upsilon _{4}(x^{k},y^{3}),
$ we solve (\ref{eqe2}) by integrating one time on $y^{3}$ and defining $%
h_{3}(x^{k},y^{3})$ for a given $h_{4}(x^{k},y^{3})$ (or, inversely, by
integrating two times on $y^{3}$ and defining $h_{4}(x^{k},y^{3})$ for a
given $h_{3}(x^{k},y^{3})$). After we determined the values $g_{i}(x^{k})$
and $h_{\ ^{0}a}(x^{k},y^{3}),$ we can compute the coefficients of
N--connection:\ The functions $w_{j}(x^{k},y^{3})$ are solutions of
algebraic equations (\ref{eqe3a}). Integrating two times on $y^{3},$ we find
$n_{j}(x^{k},y^{3}).$ The general solutions depend on integration functions
depending on coordinates $x^{k}.$ For physical considerations, we have to
consider well defined boundary conditions for such integration functions.

\section{Finsler Branes and Trapping to HL and GR}

\label{sec4} In this section, we analyze brane models when the 4--d
Horava--Lifshitz theory is embedded into 8--d Finsler spaces with
non--factorizable velocity type coordinates (experimentally, the light
velocity is finite). We shall adapt to nonholonomic and/or scale anisotropic
configurations some ideas and methods from Refs. \cite%
{vbrane,midod,gm1,gm2,gs1,gs2,singlbr} when various trapping/localizing
mechanisms for various spins $\left(0,1/2,1,2\right) $ on the 4--d brane/
observable spacetime were analyzed.

We have to consider warped Finsler geometries and analyze trapping
mechanisms because there are not experimental data for Finsler like metrics
depending on coordinates and velocities. Such dependencies can be always
derived in various isotropic and anisotropic QG models with nonlinear
dispersions. We expectations that brane trapping effects may allow us to
detect QG and LV effects experimentally even at scales much large than the
Planck one. On Finsler branes, we can consider that gravitons are allowed to
propagate in the bulk of a Finsler spacetime with dependence of
geometric/physical objects on velocity/ momentum coordinates.

\subsection{An ansatz for generating HF--brane solutions}

For constructing brane solutions in EFG, we use the ansatz for a class of
metrics which via frame transform can be parametrized in the form
\begin{eqnarray}
\mathbf{g} &=&\ \phi ^{2}(y^{5})[g_{1}(x^{k})\ e^{1}\otimes
e^{1}+g_{2}(x^{k})\ e^{2}\otimes e^{2}  \notag \\
&&+h_{3}(x^{k},v)\ \mathbf{e}^{3}\otimes \mathbf{e}^{3}+h_{4}(x^{k},v)\
\mathbf{e}^{4}\otimes \mathbf{e}^{4}]  \notag \\
&&+\left( \ ^{\ast }\mathit{l}_{P}\right) ^{2}\ [h_{5}(x^{k},v,y^{5})\
\mathbf{e}^{5}\otimes \ \mathbf{e}^{5}+h_{6}(x^{k},v,y^{5})\ \mathbf{e}%
^{6}\otimes \ \mathbf{e}^{6}]  \label{ans8d} \\
&&+\left( \ ^{\ast }\mathit{l}_{P}\right) ^{2}\ [h_{7}(x^{k},v,y^{5},y^{7})\
\mathbf{e}^{7}\otimes \ \mathbf{e}^{7}+h_{8}(x^{k},v,y^{5},y^{7})\ \mathbf{e}%
^{8}\otimes \ \mathbf{e}^{8}],  \notag
\end{eqnarray}%
\begin{eqnarray*}
 \mbox{ where } \mathbf{e}^{3} &=&dv+w_{i}dx^{i},\ \mathbf{e}^{4}=dy^{4}+n_{i}dx^{i}, \\
\mathbf{e}^{5} &=&dy^{5}+\ ^{1}w_{i}dx^{i}+\ ^{1}w_{3}dv+\ ^{1}w_{4}dy^{4},
\\
\mathbf{e}^{6} &=&dy^{6}+\ ^{1}n_{i}dx^{i}+\ ^{1}n_{3}dv+\ ^{1}n_{4}dy^{4},
\\
\mathbf{e}^{7} &=&dy^{7}+\ ^{2}w_{i}dx^{i}+\ ^{2}w_{3}dv+\ ^{2}w_{4}dy^{4}+\
^{2}w_{5}dy^{5}+\ ^{2}w_{6}dy^{6}, \\
\mathbf{e}^{8} &=&dy^{8}+\ ^{2}n_{i}dx^{i}+\ ^{2}n_{3}dv+\ ^{2}n_{4}dy^{4}+\
^{2}n_{5}dy^{5}+\ ^{2}n_{6}dy^{6},
\end{eqnarray*}%
for nontrivial N--connection coefficients
\begin{eqnarray}
N_{i}^{3} &=&w_{i}(x^{k},v),N_{i}^{4}=n_{i}(x^{k},v);  \label{ncon8d} \\
N_{i}^{5} &=&\ ^{1}w_{i}(x^{k},v,y^{5}),N_{3}^{5}=\
^{1}w_{3}(x^{k},v,y^{5}),N_{4}^{5}=\ ^{1}w_{4}(x^{k},v,y^{5});  \notag \\
N_{i}^{6} &=&\ ^{1}n_{i}(x^{k},v,y^{5});N_{3}^{6}=\
^{1}n_{3}(x^{k},v,y^{5}),N_{4}^{6}=\ ^{1}n_{4}(x^{k},v,y^{5});  \notag \\
N_{i}^{7} &=&\ ^{2}w_{i}(x^{k},v,y^{7}),N_{3}^{7}=\
^{2}w_{3}(x^{k},v,y^{7}),N_{4}^{7}=\ ^{2}w_{4}(x^{k},v,y^{7}),  \notag \\
&&N_{5}^{7}=\ ^{2}w_{3}(x^{k},v,y^{7}),N_{6}^{7}=\ ^{2}w_{4}(x^{k},v,y^{7});
\notag \\
N_{i}^{8} &=&\ ^{2}n_{i}(x^{k},v,y^{7}),N_{3}^{8}=\
^{2}n_{3}(x^{k},v,y^{7}),N_{4}^{8}=\ ^{2}n_{4}(x^{k},v,y^{7}),  \notag \\
&&N_{5}^{8}=\ ^{2}n_{3}(x^{k},v,y^{7}),N_{6}^{8}=\ ^{2}n_{4}(x^{k},v,y^{7}).
\notag
\end{eqnarray}%
The local coordinates in the above ansatz (\ref{ans8d}) are labelled in the
form $x^{i}=(x^{1},x^{2}),$ for $i,j,...=1,2;$ $y^{3}=v.$

We can include solutions of HL gravity into (\ref{ans8d}) vial polarization $%
\eta $--functions when%
\begin{eqnarray*}
g_{i}(x^{k}) &=&\eta _{i}(x^{k},v)\ ^{\circ }g_{i}(x^{k},v),\
h_{a}(x^{k},v)=\eta _{a}(x^{k},v)\ ^{\circ }h_{a}(x^{k},v), \\
N_{i}^{3}(x^{k},v) &=&\eta _{i}^{3}(x^{k},v)\ ^{\circ }w_{i}(x^{k},v),\
N_{i}^{4}(x^{k},v)=\eta _{i}^{3}(x^{k},v)\ ^{\circ }n_{i}(x^{k},v),
\end{eqnarray*}%
where "primary" data $\left[ \ ^{\circ }g_{i},\ ^{\circ }h_{a},\ ^{\circ }w_{i},\
^{\circ }n_{i}\right] $ are defined for a solution of gravitational field
equations derived from HL action (\ref{action}), and gravitational
polarizations $\left[ \eta _{i},\ \eta _{a},\ \eta _{i}^{b}\right] $ should
be defined from the condition that the "target" data $\left[ g_{i},h_{a},w_{i},n_{i}\right] $
determine solutions of the system (\ref{eqe1})--(\ref{eqe3}); the nontrivial
$[ N_{\alpha }^{5},N_{\alpha }^{6},N_{\ ^{1}\alpha }^{7},N_{\ ^{1}\alpha
}^{8},$ $h_{5},h_{6},h_{7},h_{8}]$ should be constructed as solutions of the system (\ref{eqeho}). For instance, we can take that some values with $\ ^{\circ }$ are correspondingly given by solutions on  extremal spherical and rotating black holes of Ho\v rava gravity \cite{ghodsi}  and derive generic off--diagonal generalizations, for instance, with ellipsoidal configurations like we considered in Refs. \cite{vnc2,vsingl1,vsingl2}.

The purpose of this section is to construct and analyze physical
implications of solutions of equations (\ref{ensteqcdc}) and, in particular,
(\ref{eqe1})--(\ref{eqeho}) defined by ansatz (\ref{ans8d}) with,
respectively, trivial and non--trivial N--connection coefficients (\ref%
{ncon8d}). The diagonal scenario from HF to GR is outlined in brief in
Appendix in a form very similar that for the diagonal transition from\ EFG
to GR in Ref. \cite{vbrane}. In this work, we use the canonical
d--connection instead of the Cartan d--connection.

\subsection{Finsler brane solutions}

One of the main goals of this work is to elaborate trapping scenarios for
''true'' Finsler like configurations with positively nontrivial
N--connections as solutions of nonolonomic gravitational equations (\ref%
{ensteqcdc}). The priority of such generic off--diagonal solutions is that
they allows us to distinguish the QG phenomenology and effects with LV of
(pseudo) Finsler type from that described by (pseudo) Riemannian ones
(following analysis from Introduction section the last variant is less
natural with very special types of nonlinear dispersions which must result
in vanishing N--connection structures).

\subsubsection{Separation of equations in HF models of brane gravity}

We consider an ansatz (\ref{ans8d}) multiplied to $\phi ^{2}(y^{5})$ and
with non--trivial N--connection coefficients (\ref{ncon8d}) and respective $%
\eta $--polarizations. We define the conditions when the coefficients
generate exact solutions of (\ref{ensteqcdc})for general sources of type (%
\ref{sourcb1}) and (\ref{source3}). For such ansatz, the system of equations
in HF gravity (\ref{eqe1})--(\ref{eqeho}) (we label $g_{1}=g_{2}=\epsilon
_{\pm }e^{\psi (x^{k})},$ for $\epsilon _{\pm }=\pm 1)$ transform into
\begin{eqnarray}
&&\epsilon _{\pm }\psi ^{\bullet \bullet }(x^{i})+\epsilon _{\pm }\psi
^{\prime \prime }(x^{i})=2\ ^{h}\Lambda (x^{i}),  \notag \\
&&h_{4}^{\ast }(x^{i},v)=2h_{3}(x^{i},v)\ h_{4}(x^{i},v)\ ^{v}\Lambda
(x^{i},v)/\widehat{\phi }^{\ast }(x^{i},v),  \label{4ep2b} \\
&&\partial _{y^{5}}h_{6}(u^{\alpha },y^{5})\ =2h_{5}(u^{\alpha
},y^{5})h_{6}(u^{\alpha },y^{5})\ ^{1}\Lambda u^{\alpha },y^{5})\ /\partial
_{y^{5}}\ ^{1}\widehat{\phi }(u^{\alpha },y^{5})\ ,  \notag \\
&&\partial _{y^{7}}h_{8}(u^{\ ^{1}\alpha },y^{7})=2h_{7}(u^{\ ^{1}\alpha
},y^{7})h_{6}(u^{\ ^{1}\alpha },y^{7})\ ^{2}\Lambda (u^{\ ^{1}\alpha
},y^{7})/\partial _{y^{7}}\ ^{2}\widehat{\phi }(u^{\ ^{1}\alpha },y^{7}),
\notag
\end{eqnarray}%
and the solutions for N--connection coefficients,%
\begin{eqnarray}
\beta (x^{i},v)\ w_{i}(x^{i},v)+\alpha _{i}(x^{i},v) &=&0,  \label{4ep3b} \\
\ ^{1}\beta (u^{\alpha },y^{5})\ \ ^{1}w_{\gamma }(u^{\alpha },y^{5})+\
^{1}\alpha _{\gamma }(u^{\alpha },y^{5}) &=&0,  \notag \\
\ ^{2}\beta (u^{\ ^{1}\alpha },y^{7})\ \ ^{2}w_{\ ^{1}\gamma }(u^{\
^{1}\alpha },y^{7})+\ ^{2}\alpha _{\ ^{1}\gamma }(u^{\ ^{1}\alpha },y^{7})
&=&0,  \notag
\end{eqnarray}%
\begin{eqnarray*}
n_{i}^{\ast \ast }(x^{i},v)+\gamma (x^{i},v)\ n_{i}^{\ast }(x^{i},v) &=&0, \\
\partial _{y^{5}}\partial _{y^{5}}\ ^{1}n_{\mu }(u^{\alpha },y^{5})\ +\
^{1}\gamma (u^{\alpha },y^{5})\ \partial _{y^{5}}\ \ ^{1}n_{\mu }(u^{\alpha
},y^{5})\  &=&0, \\
\partial _{y^{7}}\partial _{y^{7}}\ ^{2}n_{\mu }(u^{\ ^{1}\alpha },y^{7})\
+\ ^{2}\gamma (u^{\ ^{1}\alpha },y^{7})\ \partial _{y^{7}}\ \ ^{2}n_{\mu
}(u^{\ ^{1}\alpha },y^{7})\  &=&0,
\end{eqnarray*}%
where
\begin{eqnarray*}
\alpha _{i} &=&h_{4}^{\ast }\partial _{i}\widehat{\phi },\ \beta
=h_{4}^{\ast }\widehat{\phi }^{\ast },\ \widehat{\phi }=\ln \left| \frac{%
h_{4}^{\ast }}{\sqrt{|h_{3}h_{4}|}}\right| ,\ \gamma =\left( \frac{%
|h_{4}|^{3/2}}{|h_{3}|}\right) ^{\ast }, \\
\ ^{1}\alpha _{\mu } &=&(\partial _{y^{5}}h_{6})\partial _{\mu }\ ^{1}%
\widehat{\phi },\ \ ^{1}\beta =(\partial _{y^{5}}h_{6})(\partial _{y^{5}}\
^{1}\widehat{\phi }), \\
&&\ ^{1}\widehat{\phi }=\ln \left| \frac{\partial _{y^{5}}h_{6}\ }{\sqrt{%
|h_{5}h_{6}|}}\right| ,\ \ ^{1}\gamma =\partial _{y^{5}}\left( \frac{%
|h_{6}|^{3/2}}{|h_{5}|}\right) ,
\end{eqnarray*}%
\begin{eqnarray*}
\ ^{2}\alpha _{\ ^{1}\mu } &=&(\partial _{y^{7}}h_{8})\partial _{\ ^{1}\mu
}\ ^{2}\widehat{\phi },\ \ ^{2}\beta =(\partial _{y^{7}}h_{8})(\partial
_{y^{7}}\ ^{2}\widehat{\phi }), \\
&&\ ^{2}\widehat{\phi }=\ln \left| \frac{\partial _{y^{7}}h_{8}}{\sqrt{%
|h_{7}h_{8}|}}\right| ,\ \ ^{2}\gamma =\partial _{y^{7}}\left( \frac{%
|h_{8}|^{3/2}}{|h_{7}|}\right) ,
\end{eqnarray*}%
for $h_{3,4}^{\ast }\neq 0,\ \partial _{y^{5}}h_{6}\neq 0,\partial
_{y^{7}}h_{8}\neq 0,\ ^{h}\Lambda \neq 0,\ ^{v}\Lambda \neq 0,\ ^{b}\Lambda
\neq 0.$

\subsubsection{Exact solutions for HF brane models}

The system of partial derivative equations (\ref{4ep2b}) and (\ref{4ep3b})
is with separation of equations. It can be integrated in general form:%
\begin{eqnarray}
g_{1} &=&g_{2}=\epsilon _{\pm }e^{\psi (x^{k})},  \label{solf1} \\
h_{4} &=&\ ^{0}h_{4}(x^{k})\pm 2\int \frac{\left( \exp [2\ \widehat{\phi }%
(x^{i},v)]\right) ^{\ast }}{\ ^{v}\Lambda (x^{i},v)}dv,  \notag \\
h_{3} &=&\pm \frac{1}{4}\left[ \sqrt{|h_{4}^{\ast }(x^{i},v)|}\right]
^{2}\exp \left[ -2\ \widehat{\phi }(x^{i},v)\right] ,  \notag \end{eqnarray}
\begin{eqnarray*}
h_{6} &=&\ ^{0}h_{6}(u^{\alpha })\pm 2\int \frac{\partial _{y^{5}}\left(
\exp [2\ \ ^{1}\widehat{\phi }(u^{\alpha },y^{5})]\right) }{\ ^{1}\Lambda
(u^{\alpha },y^{5})\ }dy^{5},  \notag \\
h_{5} &=&\pm \frac{1}{4}\left[ \sqrt{|\partial _{y^{5}}h_{6}(u^{\alpha
},y^{5})|}\right] ^{2}\exp \left[ -2\ \ ^{1}\widehat{\phi }(u^{\alpha
},y^{5})\right] ,  \notag \\
h_{8} &=&\ ^{0}h_{8}(u^{\ ^{1}\alpha })\pm \int \frac{\partial
_{y^{7}}\left( \exp [2\ \ ^{2}\widehat{\phi }(u^{\ ^{1}\alpha
},y^{7})]\right) }{\ ^{2}\Lambda (u^{\ ^{1}\alpha },y^{7})\ }dy^{7},  \notag
\\
h_{7} &=&\pm \frac{1}{4}\left[ \sqrt{|\partial _{y^{7}}h_{8}(u^{\ ^{1}\alpha
},y^{7})|}\right] ^{2}\exp \left[ -2\ \ ^{2}\widehat{\phi }(u^{\ ^{1}\alpha
},y^{7})\right] ,  \notag
\end{eqnarray*}%
and, for N--connection coefficients,%
\begin{eqnarray}
w_{i} &=&-\partial _{i}\ \widehat{\phi }/\ \widehat{\phi }^{\ast },
\label{solf2} \\
n_{k} &=&\ n_{k}^{[0]}(x^{i})+\ n_{k}^{[1]}(x^{i})\int \left[ h_{3}/\left(
\sqrt{|h_{4}|}\right) ^{3}\right] dv,  \notag \\
\ ^{1}w_{\alpha } &=&-\partial _{\alpha }(\ ^{1}\widehat{\phi })/\partial
_{y^{5}}(\ ^{1}\widehat{\phi }),  \notag  \\
\ ^{1}n_{\beta } &=&\ \ ^{1}n_{\beta }^{[0]}(u^{\alpha })+\ \ ^{1}n_{\beta
}^{[1]}(u^{\alpha })\int \left[ h_{5}/\left( \sqrt{|h_{6}|}\right) ^{3}%
\right] dy^{5}, \notag  \\
\ ^{2}w_{\ ^{1}\alpha } &=&-\partial _{\ ^{1}\alpha }(\ \ ^{2}\widehat{\phi }%
)/\partial _{y^{7}}(\ ^{2}\widehat{\phi }), \notag  \\
\ ^{2}n_{\ ^{1}\beta } &=&\ \ ^{2}n_{\ ^{1}\beta }^{[0]}(u^{\ ^{1}\alpha
})+\ \ ^{2}n_{\ ^{1}\beta }^{[1]}(u^{\ ^{1}\alpha })\int \left[ h_{7}/\left(
\sqrt{|h_{8}|}\right) ^{3}\right] dy^{7}. \notag
\end{eqnarray}

The above presented classes of solutions with nonzero $h_{3}^{\ast },$ $%
h_{4}^{\ast },\partial _{y^{5}}h_{5},\partial _{y^{5}}h_{6},$ $\partial
_{y^{7}}h_{7},\partial _{y^{7}}h_{8}$ are determined by generating functions
$\widehat{\phi }(x^{i},v),\widehat{\phi }^{\ast }\neq 0;$ \newline
$\ ^{1}\ \widehat{\phi }(x^{i},y^{5}),\partial _{y^{5}}\ ^{1}\ \widehat{\phi
}\neq 0,$ $\ ^{2}\ \widehat{\phi }(x^{i},y^{5},y^{7}),$ $\partial _{y^{7}}\
^{2}\ \widehat{\phi }\neq 0,$ and integration functions $\
n_{k}^{[0]}(x^{i}),$ $\ n_{k}^{[1]}(x^{i}),\ \ \ ^{1}n_{\beta
}^{[0]}(u^{\alpha }),\ ^{1}n_{\beta }^{[1]}(u^{\alpha }),\ \ ^{2}n_{\
^{1}\beta }^{[0]}(u^{\ ^{1}\alpha }),\ \ ^{2}n_{\ ^{1}\beta }^{[1]}(u^{\
^{1}\alpha }).$ In order to construct explicit solutions, we have to chose
and/or fix such functions following additional assumptions on symmetry of
solutions, boundary conditions etc.

The coefficients (\ref{solf1}) and (\ref{solf2}) can be additionally
constrained if we wont to construct solutions for the Levi--Civita
connection on $T\mathbf{V.}$ By straightforward computations (see details in %
\cite{vex1,vex2,vex3}), we can verify that all torsion coefficients (\ref%
{dtors}) vanish if
\begin{eqnarray*}
w_{i}^{\ast } &=&\mathbf{e}_{i}\ln |h_{4}|,\ \mathbf{e}_{k}w_{i}=\mathbf{e}%
_{i}w_{k},\ n_{i}^{\ast }=0, \partial _{i}n_{k}=\partial _{k}n_{i}, \\
\partial _{y^{5}}(\ ^{1}w_{\alpha }) &=&\ ^{1}\mathbf{e}_{\alpha }\ln
|h_{6}|,\ \ ^{1}\mathbf{e}_{\alpha }\ ^{1}w_{\beta }=\ ^{1}\mathbf{e}_{\beta
}\ ^{1}w_{\alpha }, \\
&& \partial _{y^{5}}(\ ^{1}n_{\alpha })=0, \partial _{\alpha }\ ^{1}n_{\beta
}=\partial _{\beta }\ ^{1}n_{\alpha },
\end{eqnarray*}
\begin{eqnarray*}
\partial _{y^{7}}(\ ^{2}w_{\ ^{1}\alpha }) &=&\ ^{2}\mathbf{e}_{\ ^{1}\alpha
}\ln |h_{8}|,\ \ ^{2}\mathbf{e}_{\ ^{1}\alpha }\ ^{2}w_{\ ^{1}\beta }=\ ^{2}%
\mathbf{e}_{\ ^{1}\beta }\ ^{2}w_{\ ^{1}\alpha }, \\
&& \partial _{y^{7}}(\ ^{2}n_{\ ^{1}\alpha })=0,\ \partial _{\ ^{1}\alpha }\
^{2}n_{\ ^{1}\beta }=\partial _{\ ^{1}\beta }\ ^{2}n_{\ ^{1}\alpha }.
\end{eqnarray*}%
Such conditions can be satisfied by imposing certain constraints on the
considered classes of generating and integration functions. This class of
generic off--diagonal solutions are important if we wont to construct
trapping configurations from the HF brane gravity to GR on $\mathbf{V,}$
when the conditions (\ref{grcond}) are imposed.

\subsubsection{Remarks on (non) diagonal HF brane solutions on $T\mathbf{V}$}

The above solutions in HF gravity are still very general. It is not clear
what physical meaning they may have and we must impose additional
restrictions on some coefficients of metrics and sources in order to
construct in explicit form certain Finsler brane configurations resulting in
HL, or GR, theories and model a trapping mechanism with generic
off--diagonal metrics.

There is a class of sources in HF gravity when for trivial N--connection
coefficients (i.e. for zero values (\ref{ncon8d})) the sources $\widetilde{%
\mathbf{\Upsilon }}_{\ \ ^{2}\delta }^{\ ^{2}\beta }$ (\ref{sourcb})
transform into data (which in the diagonal limit we get sources for the
gravitational equations for the Levi--Civita connection labeled with a left
low bar) $\ _{\shortmid }\Upsilon _{\ \ ^{2}\delta }^{\ ^{2}\beta }$ (\ref%
{source3}), with nontrivial limits for $\ _{\shortmid }\Upsilon _{\ \delta
}^{\beta }=\Lambda -M^{-(m+2)}\overline{K}_{1}(y^{5})$ and $\ _{\shortmid
}\Upsilon _{\ 5}^{5}=\ _{\shortmid }\Upsilon _{\ 6}^{6}=\Lambda -M^{-(m+2)}%
\overline{K}_{2}(y^{5}),$ being preserved certain conditions of type (\ref%
{sourcb1}). The generating functions are taken in the form when
\begin{eqnarray*}
h_{5}(x^{i},y^{5}) &=&\ ^{\ast }\mathit{l}_{P}\frac{\overline{h}(y^{5})}{%
\phi ^{2}(y^{5})}\ ^{q}h_{5}(x^{i},y^{5}),\\
 h_{6}(x^{i},y^{5})&=&\ ^{\ast }%
\mathit{l}_{P}\frac{\overline{h}(y^{5})}{\phi ^{2}(y^{5})}\
^{q}h_{6}(x^{i},y^{5}), \\
h_{7}(x^{i},y^{5},y^{7}) &=&\ ^{\ast }\mathit{l}_{P}\frac{\overline{h}(y^{5})%
}{\phi ^{2}(y^{5})}\ ^{q}h_{7}(x^{i},y^{5},y^{7}), \\
h_{8}(x^{i},y^{5},y^{7}) &=&\ ^{\ast }\mathit{l}_{P}\frac{\overline{h}(y^{5})%
}{\phi ^{2}(y^{5})}\ ^{q}h_{8}(x^{i},y^{5},y^{7}),
\end{eqnarray*}%
where the generating functions are parametrized in such a form that $\phi
^{2}(y^{5})$ and $h_{5}(y^{5})$ are those for diagonal metrics and $\
^{q}h_{5},\ ^{q}h_{6},\ ^{q}h_{7},$ $\ ^{q}h_{8}$ are computed following
formulas (\ref{solf1}) and (\ref{solf2}). This class of off--diagonal
metrics are parametrized in the form
\begin{eqnarray}
\mathbf{g} &=&g_{1}dx^{1}\otimes dx^{1}+g_{2}dx^{2}\otimes dx^{2}+h_{3}%
\mathbf{e}^{3}{\otimes }\mathbf{e}^{3}\ +h_{4}\mathbf{e}^{4}{\otimes }%
\mathbf{e}^{4}\ +  \label{fbr} \\
&&\left( \ ^{\ast }\mathit{l}_{P}\right) ^{2}\frac{\overline{h}}{\phi ^{2}}%
[\ \ ^{q}h_{5}\mathbf{e}^{5}\otimes \ \mathbf{e}^{5}+\ ^{q}h_{6}\mathbf{e}%
^{6}\otimes \ \mathbf{e}^{6}+\ ^{q}h_{7}\mathbf{e}^{7}\otimes \ \mathbf{e}%
^{7}+\ ^{q}h_{8}\mathbf{e}^{8}\otimes \ \mathbf{e}^{8}],  \notag
\end{eqnarray}%
where
\begin{eqnarray}
\ \mathbf{e}^{3} &=&dy^{3}+w_{i}dx^{i},\mathbf{e}^{4}=dy^{4}+n_{i}dx^{i},
\label{ncfbr} \\
\ \mathbf{e}^{5} &=&dy^{5}+\ ^{1}w_{i}dx^{i},\mathbf{e}^{6}=dy^{6}+\
^{1}n_{i}dx^{i},  \notag \\
\mathbf{e}^{7} &=&dy^{7}+\ ^{2}w_{i}dx^{i},\mathbf{e}^{8}=dy^{8}+\
^{2}n_{i}dx^{i}.  \notag
\end{eqnarray}%
Such off--diagonal parameterizations of metrics where considered in \cite%
{vbrane} but the coefficients of the metric and N--connection where computed
there for a different d--connection (for the Cartan d--connection).

Any solution of type (\ref{fbr}) describes an off--diagonal canonical
nonholonomic trapping for 8--d (respectively, for corresponding classes of
generating and integration functions, 5--, 6--, 7--d) to 4--d modifications
of HL and/or GR with some corrections depending on bulk Finsler
''fluctuations'' and LV effects. There is a class of sources when for
vanishing N--connection coefficients (\ref{ncfbr}) we get diagonal metrics
of type (\ref{ans8d}), considered in Appendix, but multiplied to a conformal
factor $\phi ^{2}(y^{5})$ when the $h$--coefficients are solutions of
equations of type (\ref{4ep2b}).

With respect to local coordinate cobase $du^{\ ^{2}\alpha
}=(dx^{i},dy^{a},dy^{\ ^{1}a},dy^{\ ^{2}a})$ a solution (\ref{fbr}) is
parametrized by an off--diagonal matrix $g_{\ ^{2}\alpha \ ^{2}\beta }=$%
%{\tiny
\begin{equation*}
\left[
\begin{array}{cccccccc}
B_{11} & B_{12} & w_{1}h_{3} & n_{1}h_{4} & \ ^{1}w_{1}h_{5} & \
^{1}n_{1}h_{6} & \ ^{2}w_{1}h_{7} & \ ^{2}n_{1}h_{8} \\
B_{21} & B_{22} & w_{2}h_{3} & n_{2}h_{4} & \ ^{1}w_{2}h_{5} & \
^{1}n_{2}h_{6} & \ ^{2}w_{2}h_{7} & \ ^{2}n_{2}h_{8} \\
w_{1}h_{3} & w_{2}h_{3} & h_{3} & 0 & 0 & 0 & 0 & 0 \\
n_{1}h_{4} & n_{2}h_{4} & 0 & h_{4} & 0 & 0 & 0 & 0 \\
\ ^{1}w_{1}h_{5} & ^{1}w_{2}h_{5} & 0 & 0 & h_{5} & 0 & 0 & 0 \\
\ ^{1}n_{1}h_{6} & \ ^{1}n_{2}h_{6} & 0 & 0 & 0 & h_{6} & 0 & 0 \\
\ ^{2}w_{1}h_{7} & \ ^{2}w_{2}h_{7} & 0 & 0 & 0 & 0 & h_{7} & 0 \\
\ ^{2}n_{1}h_{8} & \ ^{2}n_{2}h_{8} & 0 & 0 & 0 & 0 & 0 & h_{8}%
\end{array}%
\right]
\end{equation*}%
%
%}
where possible observable Finsler brane and LV contributions are
distinguished by terms proportional to $\left( \ ^{\ast }\mathit{l}%
_{P}\right) ^{2}$ in
\begin{eqnarray*}
B_{11} &=&g_{1}+w_{1}^{2}h_{3}+n_{1}^{2}h_{4}+\left( \ ^{\ast }\mathit{l}%
_{P}\right) ^{2}\frac{\overline{h}}{\phi ^{2}}\times \\
&&\left[ (\ ^{1}w_{1})^{2}\ ^{q}h_{5}+(\ ^{1}n_{1})^{2}\ ^{q}h_{6}+(\
^{2}w_{1})^{2}\ ^{q}h_{7}+(\ ^{2}n_{1})^{2}\ ^{q}h_{8}\right] , \\
B_{12} &=&B_{21}=w_{1}w_{2}h_{3}+n_{1}n_{2}h_{4}+\left( \ ^{\ast }\mathit{l}%
_{P}\right) ^{2}\frac{\overline{h}}{\phi ^{2}}\times \\
&&\left[ \ ^{1}w_{1}\ ^{1}w_{2}\ ^{q}h_{5}+\ ^{1}n_{1}\ ^{1}n_{2}\
^{q}h_{6}+\ ^{2}w_{1}\ ^{2}w_{2}\ ^{q}h_{7}+\ ^{2}n_{1}\ ^{2}n_{2}\ ^{q}h_{8}%
\right] , \\
B_{22} &=&g_{2}+w_{2}^{2}h_{3}+n_{2}^{2}h_{4}+\left( \ ^{\ast }\mathit{l}%
_{P}\right) ^{2}\frac{\overline{h}}{\phi ^{2}}\times \\
&&\left[ (\ ^{1}w_{2})^{2}\ ^{q}h_{5}+(\ ^{1}n_{2})^{2}\ ^{q}h_{6}+(\
^{2}w_{2})^{2}\ ^{q}h_{7}+(\ ^{2}n_{2})^{2}\ ^{q}h_{8}\right] .
\end{eqnarray*}%
As a matter of principle, it is possible to distinguish \ experimentally
some generic off--diagonal metrics in Finsler geometry from certain diagonal
configurations of type (\ref{ansdiag8d}) with Levi--Civita connection on $T%
\mathbf{V}.$

Let us discuss and compare the above Finsler brane off--diagonal solutions
constructed above and those studied in Ref. \cite{vbrane}. The formulas for
coefficients (\ref{solf1}) and (\ref{solf2}) computed (in this work) for the
canonical d--connection describe an off--diagonal brane extension to a
Finsler spacetime from the HL gravity with scaling anisotropy (we proved
that there is a trapping mechanism encoded into such solutions relating HF
and HL gravity models). For Finsler branes induced from a QG models with LV,
based on "nonrenormalizable" GR (which we studied in the just mentioned our
paper) the trapping scenario was modelled by the Cartan d--connection on a
Finsler brane and the resulting configuration was certain one in "locally
isotropic" GR theory.

\section{Discussion and Conclusions}

\label{sec5} Summarizing the results of this paper (see also a series of partner works \cite{vcosm,vbrane,vnc3,vcrit}), we conclude that there are at least nine substantial arguments to consider that Finsler geometry and related geometric methods are of crucial importance in modern classical and quantum gravity, particle physics, cosmology and modifications:
\begin{enumerate}
\item The bulk of models of quantum gravity (QG) and related phenomenology
are with Lorentz violation (LV) being characterized by corresponding
modified dispersion relations (MDR). In its turn, such a MDR determines
naturally a fundamental/generating Finsler function on (co) tangent space.
This QG--LV--MDR--Finsler geometry scheme works for various models of QG
with limits to, or warped/trapped, configurations derived in (super) string/
brane/ noncommutative/ analogous gravity/ gauge gravity etc theories. The Ho%
\v{r}ava--Lifshitz (HL) theory with scaling anisotropy and MDR can be
included into such a generalized Finsler gravity scheme.

\item Locally anisotropic structures and Finsler geometries are considered in analogous gravity, geometric mechanics and various models of condensed
matter physics; certain important ideas and methods from physics of phase transitions are exploited in modern QG and phenomenology.

\item The ideas on restricted special relativity, scenarios with LV, modified/ generalized Lorentz symmetries have straightforward relations to some special models of Finsler geometry, anisotropic symmetries and corresponding local/global transformation laws.

\item There are certain ideas and explicit constructions \ suggesting that various problems related to dark energy and dark matter physics, accelerating Universes, anisotropies etc can be cured by modifying the pseudo--Riemannian/ Lorentzian spacetime paradigm to (pseudo) Finsler spacetimes and generalizations.

\item Finsler like geometries are ''canonically'' generated/ induced as exact solutions of nonholonomic Ricci flows of (pseudo) Riemannian metrics, and
for various evolution scenarios with noncommutative and/or nonsymmetric metrics, gravitational diffusion and stochastic processes, fractional derivatives and/or fractional dimensions, memory and self--organization etc.

\item Noncommutative generalizations of gravity theories can be modelled equivalently as complex Finsler like geometries.

\item Finsler configurations can be derived as  exact solutions of gravitational field equations in GR, string, brane, gauge gravity theories.

\item Finsler geometry methods happen to be very efficient in elaborating a new geometric method (the so--called anholonomic deformation method) of
constructing exact solutions in gravity, even for the general relativity (GR) theory. Such an approach allows us to generate very general classes of
exact solutions of Einstein equations and generalizations (with generic off--diagonal metrics, linear and nonlinear connections and nonholonomic frame coefficients depending generally on all coordinates etc). Constraining nonholonomically certain general integral varieties of solutions with generalized connections, we obtain subvarieties for the Levi--Civita connections in GR. Such a method of constructing exact solutions can be
applied in HL gravity.

\item Re--writing the Einstein gravity and/or certain generalizations in canonical Finsler variables, and then using almost K\"{a}hler equivalents,
we can quantize various types of gravity theories using methods of deformation quantization, A--brane approach, nonholonomic canonical quantization etc. It seems that it is possible to renormalize gravity using
the so--called bi--connection formalism and/or HL approach.
\end{enumerate}

Following the above mentioned reasons, we consider that HL\ gravity should be extended in a form encoding also the physics of MDR, nonholonomic configurations and anisotropic configurations. In explicit form, we
elaborated a model of Ho\v{r}ava--Finsler (HF) gravity following generalized relativity principles \cite{vcosm,vcrit,vrev1,vsgg}. We used metric
compatible distinguished connections from Finsler geometry which allows us to formulate and study classical and quantum models following standard approaches with spinors, Dirac operators and vielbeins, metrics and connections as in GR but adapted to nonholonomic structures, in our
approach, to nonlinear connections (N--connection).

The HF gravity theory is canonically formulated on tangent bundle. From a formal point of view, it is generally integrable and can be quantized/renormalized following standard methods. There are many open issues regarding HL and HF gravity models. Here, we emphasize the following. A sensible problem to be solved is that why classical limits do not contain
anisotropies and dependencies on velocity/momentum type coordinates. In explicit form, we can apply certain ideas and methods from brane gravity which was studied intensively during last twelve years beginning Gogberashvili and Rundal--Sundrum works. Nevertheless, for Finsler branes
and HF gravity, such constructions can not be applied in a straightforward form. Possible warping, trapping, compactification etc scenarios for Finsler
spaces should encode, in general, a nontrivial N--connection structure. Technically, to construct such HF brane exact solutions with generic off--diagonal metrics is a very difficult task. One of the main results of
this work is that we were able to solve and analyze such off--diagonal locally anisotropic trapping scenario from HF to HL and/or GR theories. Such a nonholonomic gravitational dynamics encode also in general form various
types of MDR, parametric dependencies, possible generalized symmetries etc.\

The length of this paper does not allow us to address the question of stability of Finsler brane solutions. In general, stable configurations can be constructed for diagonal solutions which survive for nonholonomically constrained off--diagonal ones (proofs are similar to those for extra
dimensional brane solutions; we plan to study the problem in details in our further works). Hopeful, future work will concern various topics from QG
with LV and Finser geometry methods and possible applications in modern cosmology and astrophysics.

\vskip5pt

\textbf{Acknowledgements: }  I'm grateful for support, hospitality and/or  important discussions on generalized/modified Finsler gravity to  G. Calcagni, E. Elizalde, M. Francavigla, C. L\"{a}mmerzahl, N. Mavromatos, S. Odintsov,  D. Pavlov, V. Perlick, E. Radu, S. Sarkar, F. P. Schuller,  L. Sindoni and P. Stavrinos.  I thank E. Hatefi, A. Kobakhidze, F. Mercati and D. Orlando for some critical remarks and pointing additional references related to existing problems,  present status and further developments of HL models. The research for this paper was partially supported by Romanian Government via  Program IDEI, PN-II-ID-PCE-2011-3-0256.

\appendix

\setcounter{equation}{0} \renewcommand{\theequation}
{A.\arabic{equation}} \setcounter{subsection}{0}
\renewcommand{\thesubsection}
{A.\arabic{subsection}}

\section{Holonomic Configurations for HF--branes}

A trapping scenario from HF to GR with diagonal metrics can be constructed
for a simplified ansatz (\ref{ans8d}) with zero N--connection coefficients (%
\ref{ncon8d}) when $h_{5},h_{7},h_{8}=const$ and data $\left[ g_{i},h_{a}%
\right] $ define a trivial solution in GR and the local signature for
metrics os of type $(+,-,-,...-).$\footnote{%
The aholonomic deformation method allows us to construct exact solutions
with any signature we consider physically important; for Finsler brane
configurations, we adapt the constructions form \cite%
{midod,gm1,gm2,gs1,gs2,singlbr} in nonolonomic form as in \cite{vbrane}.}
Such metrics can be written in the form
\begin{eqnarray}
\mathbf{g} &=&\ \phi ^{2}(y^{5})\eta _{\alpha \beta }du^{\alpha }\otimes
du^{\beta }-  \label{ansdiag8d} \\
&&\left( \mathit{l}_{P}\right) ^{2}\overline{h}(y^{5})[\ dy^{5}\otimes \
dy^{5}+dy^{6}\otimes \ dy^{6}\pm dy^{7}\otimes \ dy^{7}\pm dy^{8}\otimes \
dy^{8}],  \notag
\end{eqnarray}%
where $\eta _{\alpha \beta }=diag[1,-1,-1-,1]$ and $\alpha ,\beta
,...=1,2,3,4;$ extra dimension indices will be considered of type $\
^{1}\alpha =(\alpha ,5,6)$ and $\ ^{2}\alpha =(\ ^{1}\alpha ,7,8).$ Indices
of type $\ ^{2}\alpha ,\ ^{2}\beta ,...$ will run values $1,2,3,4,5,...,m$
where $m\geq 2.$ The fiber coordinates $y^{5},y^{6},y^{7},y^{8}$ are
velocity type. We analyze here a toy model when sources are defined by a
cosmological constant $\Lambda $ and nonzero components of stress--energy
tensor,
\begin{equation}
\ \Upsilon _{\ \delta }^{\beta }=\Lambda -M^{-(m+2)}\overline{K}%
_{1}(y^{5}),\ \Upsilon _{\ 5}^{5}=\ \Upsilon _{\ 6}^{6}=\Lambda -M^{-(m+2)}%
\overline{K}_{2}(y^{5}),  \label{source3}
\end{equation}%
for a fundamental mass scale $M$ on $T\mathbf{V},$ $\dim T\mathbf{V}=8.$

A metric (\ref{ansdiag8d}) generates a solution of gravitational field
equation in HF gravity if
\begin{equation}
\phi ^{2}(y^{5})=\frac{3\epsilon ^{2}+a(y^{5})^{2}}{3\epsilon
^{2}+(y^{5})^{2}}\mbox{ and }\ ^{\ast }\mathit{l}_{P}\sqrt{|\overline{h}%
(y^{5})|}=\frac{9\epsilon ^{4}}{\left[ 3\epsilon ^{2}+(y^{5})^{2}\right] ^{2}%
}.  \label{cond1}
\end{equation}%
In the above formulas we consider $a$ as an integration constant and the
width of brane is $\epsilon ,$ with some fixed integration parameters when $%
\frac{\partial ^{2}\phi }{\partial (y^{5})^{2}}\mid _{y^{5}=\epsilon }=0$
and $\ ^{\ast }\mathit{l}_{P}\sqrt{|\overline{h}(y^{5})|}\mid _{y^{5}=0}=1;$
this states the conditions that on diagonal branes the Minkowski metric on $T%
\mathbf{V}$ is 6--d, or 8--d.

The sources (\ref{source3}) are compatible with the field equations if
\begin{eqnarray}
\overline{K}_{1}(y^{5})M^{-(m+2)} &=&\Lambda +\left[ 3\epsilon
^{2}+(y^{5})^{2}\right] ^{-2}[\frac{2\ ^{0}\phi m(\ ^{0}\phi (m+2)-3)}{%
3\epsilon ^{2}}(y^{5})^{4}+  \notag \\
&&2(-2\ ^{0}\phi (m^{2}+2m+6)+3(m+3)(1+\ ^{0}\phi ^{2}))(y^{5})^{2}  \notag \\
&&-6\epsilon ^{2}m(m-3\ ^{0}\phi +2)]  \label{cond2}
\end{eqnarray}
\begin{eqnarray}
\overline{K}_{2}(y^{5})M^{-(m+2)} &=&\Lambda +\left[ 3\epsilon
^{2}+(y^{5})^{2}\right] ^{-2}[\frac{2\ ^{0}\phi (m-1)(\ ^{0}\phi (m+2)-4)}{%
3\epsilon ^{2}}\times  \notag \\
&&(y^{5})^{4} +4(-\ ^{0}\phi (m^{2}+m+10)+2(m+2)(1+\ ^{0}\phi ^{2})),  \notag
\\
&&\times (y^{5})^{2} -6\epsilon ^{2}(m-1)(m-4\ ^{0}\phi +2)].  \notag
\end{eqnarray}
For Finsler branes, the width $\epsilon ^{2}=40M^{4}/3\Lambda $ is with
extra velocity type coordinates and certain constants are related to $\
^{\ast }\mathit{l}_{P}.$

For the considered diagonal ansatz, the coefficients of the canonical
d--connection are the same as for the for the Levi--Civita connection when
\begin{equation}
\nabla _{\ ^{2}\alpha }\ \Upsilon _{\ \quad }^{\ ^{2}\alpha \ ^{2}\beta }=(%
\sqrt{|\ ^{F}\mathbf{g}|})^{-1}\mathbf{e}_{\ ^{2}\alpha }(\sqrt{|\ ^{F}%
\mathbf{g}|}\ \Upsilon _{\ \quad }^{\ ^{2}\alpha \ ^{2}\beta })+\ \Gamma _{\
\ ^{2}\alpha \ ^{2}\gamma }^{\ ^{2}\beta }\ \Upsilon _{\ \quad }^{\
^{2}\alpha \ ^{2}\gamma }=0,  \label{cond3}
\end{equation}%
which for the conditions (\ref{cond1}) and (\ref{cond2}) such a conservation
law is satisfied if
\begin{equation}
\frac{\partial \overline{K}_{1}}{\partial (y^{5})}=4\left( \overline{K}_{2}-%
\overline{K}_{1}\right) \frac{\partial \ln |\phi |}{\partial (y^{5})}.
\label{cond3a}
\end{equation}

We constructed a metric (\ref{ansdiag8d}) with coefficients subjected to
conditions (\ref{cond1}) -- (\ref{cond3a}). Such a solution defines trapping
solutions containing ''diagonal'' extensions of GR to a 8--d $T\mathbf{V}$
and/or possible restrictions to 6--d and 7--d (to consider HF
configurations, we have to include off--diagonal interactions). Such
solutions provide also mechanisms of corresponding gravitational trapping
for fields of spins $0,1/2,1,2$ (proofs are very similar to those presented in Refs. \cite{midod,gm1,gm2,gs1,gs2,singlbr}).


\begin{thebibliography}{99}
\bibitem{horava1} P. Ho\v{r}ava, Membranes at quantum criticality, JHEP, \textbf{020} (2009) 0903

\bibitem{horava2} P. Ho\v rava, Quantum gravity at a Lifshitz point, Phys. Rev. \textbf{D79} (2009) 084008

\bibitem{horava3} P. Ho\v rava, Spectral dimension of the Universe in
quantum gravity at a Lifshitz point, Phys. Rev. Lett. \textbf{102} (2009)
161301

\bibitem{orlando1} D. Orlando and S. Reffert, On the renormalizability of
Horava--Lifshitz--type gravities, Class. Quant. Grav. \textbf{\ 26 } (2009) 155021

\bibitem{orlando2} D. Orlando and S. Reffert, On the Perturbative Expansion around a Lifshitz Point, Phys. Lett. \textbf{ B 683 }  (2010) 62--68

\bibitem{kost4} V. A. Kostelecky and J. D. Tasson, Matter--gravity couplings and Lorenz violation, Phys. Rev. \textbf{D 83}  (2011) 016012

\bibitem{xiao} Zhi Xiao and Bo--Qiang Ma, Constraints on Lorentz invariance violation from gamma--ray burst GRB090510, Phys. Rev. D \textbf{80} (2009) 116005

\bibitem{liberati} S. Liberati and L. Maccione, Lorentz violation; motivation and new constraints, Ann. Rev. Nucl. Part. Sci. \textbf{\ 59 } (2009) 245--267

\bibitem{carroll} S. M. Carroll, J. A. Harvey, V. A. Kostelecky, C. D. Lane and T. Okamoto, Noncommutative field theory and Lorentz violation, Phys. Rev. Lett. \textbf{87} (2001) 141601

\bibitem{burgess} C. P. Burgess, J. Cline, E. Filotas, J. Matias and G. D.
Moore, Loop--Generated bounds on changes to the graviton dispersion
relation, JHEP \textbf{0203 }(2002) 043

\bibitem{barcello} C. Barcelo, S. Liberati and M. Visser, Analogue gravity, Living Rev. Rel. \textbf{8} (2005) 12

\bibitem{kobakh} A. Kobakhidze, On the infrared limit of Ho\v rava's gravity with the global Hamiltonian constraint, Phys. Rev. D \textbf{\ 82 } (2010) 064011

\bibitem{blas} D. Blas, O. Pujolas and S. Sibiryakov, On the extra mode and inconsistency of Ho\v rava Gravity, JHEP \textbf{\ 0910 } (2009) 029

\bibitem{odintsov} E. Elizalde, S. Nojiri, S. D. Odintso, D. Saez--Gomez, Unifying inflation with dark energiy in modified F(R) Horava-Lifshitz gravity, Eur. Phys. J. \textbf{C 70}\  (2010) 351--361

\bibitem{carloni} S. Carloni, E. Elizalde and P. J. Silva, Matter couplings in Horava--Lifshitz and their cosmological applications, Class. Quant. Grav. \textbf{ 28 } (2011) 195002

\bibitem{mercati} G. Amelino--Camelia, L. Gualtieri and F. Mercati, Threshold anomalies in Ho\v rava--Lifshitz--type theories, Phys. Lett. B \textbf{\ 686 } (2010) 283--287

\bibitem{mavromatos} N. E. Mavromatos, Lorentz Invariance Violation from String Theory, arXiv: 0708.2250

\bibitem{mavromatos1} J. Ellis and. N. E. Mavromatos, Probes of Lorentz Violation, arXiv: 1111.1178

\bibitem{mavromatos2} N. E. Mavromatos, S. Sarkar and A. Vergou, Stringy Space--Time Foam, Finsler--like Metrics and Dark Matter Relics, Phys. Lett. \textbf{ B 696 } (2011) 300-304

\bibitem{mignemi} S. Mignemi, Doubly special relativity and Finsler geometry, Phys. Rev. D \textbf{\ 76 } (2007) 047702

\bibitem{girelli} F. Girelli, S. Liberati and L. Sindoni, Phenomenology of
quantum gravity and Finsler geometry, Phys. Rev. D \textbf{\ 75 } (2007) 064015

\bibitem{gibbons} G. W. Gibbons, J. Gomis and C. N. Pope, General very special relativity is Finsler geometry, Phys. Rev. D \textbf{76} (2007) 081701

\bibitem{sindoni} L. Sindoni, The Higgs mechanism in Finsler spacetimes, Phys. Rev. D \textbf{77} (2008) 124009

\bibitem{lammer} C. L\"{a}mmerzahl, D. Lorek and H. Dittus, Confronting Finsler space--time with experiment,\ Gen. Rel. Grav. \textbf{\ 41 } (2009) 1345-1353

\bibitem{stavrinos} A. P. Kouretsis, M. Stathakopoulos and P. C. Stavrinos, The general very special relativity in Finsler cosmology, Phys. Rev. D
\textbf{79} (2009) 104011

\bibitem{stavr0} A. P. Kouretsis, M. Stathakopoulos and P. C. Stavrinos,
Imperfect fluids, Lorentz violations and Finsler cosmology, Phys. Rev. D
\textbf{\ 82 } (2010) 064035

\bibitem{visser1} J.\ Skakala and M. Visser, Pseudo--Finslerian spacetimes
and multi--refrigerence, Int. J. Mod. Phys. D \textbf{\ 19 } (2010) 119--1146

\bibitem{visser2} S. Weinfurther, T. P. Sotriou and M. Visser, Projectable
Ho\v rava--Lifshitz gravity in a nutshell, J. Phys. Conf. Ser. \textbf{222} (2010) 012054

\bibitem{yang} H. S. Yang, Emergent spacetime and the origin of gravity,
JHEP \textbf{5} (2009) 012

\bibitem{calcagni} G. Calcagni, Detailed balance in Horava--Lifshitz
gravity, Phys.\ Rev. D \textbf{81} (2010) 044006

\bibitem{schuller} D. Raetzel, S. Rivera and F. P. Schuller, Geometry of
Physical Dispersion Relations, Phys. Rev. \textbf{ D 83 } (2011) 0444047

\bibitem{sindoni1} L. Sindoni, A Note on Particle Kinematics in Ho\v
rava--Lifshitz Scenarios, arXiv: 0910.1329

\bibitem{vrev1} S. Vacaru, Finsler and Lagrange Geometries in Einstein and String Gravity, Int. J. Geom. Methods. Mod. Phys. \textbf{5} (2008) 473-511

\bibitem{vcosm} S. Vacaru, Principles of Einstein--Finsler Gravity and
Perspectives in Modern Cosmology, arXiv: 1004.3007

\bibitem{vbrane} S. Vacaru, Finsler Branes and Quantum Gravity Phenomenology with Lorentz Symmetry Violations,  Class. Quant. Grav. \textbf{ 28} (2011) 215001

\bibitem{vsgg} S. Vacaru, P. Stavrinos, E. Gaburov and D. Gon\c{t}a, \textit{%
Clifford and Riemann- Finsler Structures in Geometric Mechanics and Gravity,}%
\ Selected Works, Differential Geometry -- Dynamical Systems, Monograph
\textbf{\ 7 } (Geometry Balkan Press, 2006);\newline
www.mathem.pub.ro/dgds/mono/va-t.pdf and arXiv: gr-qc/0508023

\bibitem{vcrit} S. Vacaru, Critical remarks on Finsler modifications of
gravity and cosmology by Zhe Chang and Xin Li, Phys. Lett. B \textbf{\ 690 } (2010) 224-228

\bibitem{ma} R. Miron and M. Anastasiei, The Geometry of Lagrange Spaces:\ Theory and Applications, FTPH no. \textbf{59} (Kluwer Academic Publishers, Dordrecht, Boston, London, 1994)

\bibitem{vstr1} S. Vacaru, Locally anisotropic gravity and strings, Ann.
Phys. (NY), \textbf{\ 256 } (1997) 39-61

\bibitem{vstr2} S. Vacaru, Superstrings in higher order extensions of
Finsler superspaces, Nucl. Phys. B, \textbf{434} (1997) 590 -656

\bibitem{vsingl1} S. Vacaru and D. Singleton, Warped solitonic deformations and propagation of black holes in 5D vacuum gravity, Class. Quant. Grav. \textbf{19} (2002) 3583-3602

\bibitem{vsingl2} S. Vacaru and D. Singleton, Warped, anisotropic wormhole soliton configurations in vacuum 5D gravity, Class. Quant. Grav. \textbf{19} (2002), 2793-2811

\bibitem{vnc1} S. Vacaru, Gauge and Einstein gravity from non--Abelian gauge models on noncommutative spaces, Phys. Lett. B \textbf{498} (2001) 74-82

\bibitem{vnc2} S. Vacaru, Exact solutions with noncommutative symmetries in Einstein and gauge gravity, J. Math. Phys. \textbf{46} (2005) 042503

\bibitem{vnc3} S. Vacaru, Finsler black holes induced by noncommutative
anholonomic distributions in Einstein gravity, Class. Quant. Grav. \textbf{27 } (2010) 105003

\bibitem{vrf1} S. Vacaru, Nonholonomic Ricci flows: II. Evolution equations
and dynamics, J. Math. Phys. \textbf{49} (2008) 043504

\bibitem{vrf2} S. Vacaru, Ricci flows and solitonic pp--waves, Int. J. Mod. Phys. \textbf{A 21} (2006) 4899-4912

\bibitem{vrf3} S. Vacaru, Nonholonomic Ricci flows, exact solutions in
gravity, and symmetric and nonsymmetric metrics, Int. J. Theor. Phys.
\textbf{48} (2009) 579-606

\bibitem{vrf4} S. Vacaru, Spectral functionals, nonholonomic Dirac
operators, and noncommutative Ricci flows, J. Math. Phys.\textbf{\ 50}
(2009) 073503

\bibitem{vex1} S. Vacaru, Parametric nonholonomic frame transforms and exact solutions in gravity, Int. J. Geom. Meth. Mod. Phys.  \textbf{4} (2007) 1285-1334

\bibitem{vex2} S. Vacaru, On general solutions in Einstein and high
dimensional gravity, Int. J. Theor. Phys. \textbf{49} (2010) 884-913

\bibitem{vex3} S. Vacaru, On general solutions in Einstein gravity,
Int. J. Geom. Meth. Mod. Phys. \textbf{8} (2011) 9--21

\bibitem{vqgr1} S. Vacaru, Deformation quantization of almost K\"{a}hler models and Lagrange--Finsler spaces, J. Math. Phys. \textbf{48} (2007) 123509

\bibitem{vqgr2} S. Vacaru, Deformation quantization of nonholonomic almost K%
\"{a}hler models and Einstein gravity, Phys. Lett. A \textbf{ 372 }  (2008) 2949-2955

\bibitem{vqgr3} M. Anastasiei and S. Vacaru, Fedosov quantization of
Lagrange--Finsler and Hamilton--Cartan spaces and Einstein gravity lifts on (co) tangent bundles, J. Math. Phys. \textbf{50} (2009) 013510

\bibitem{vqgr4} S. Vacaru, Branes and quantization for an A--model
complexification of Einstein gravity in almost K\"{a}hler variables, Int. J.
Geom. Meth. Mod. Phys. \textbf{6} (2009) 873-909

\bibitem{vqgr5} S. Vacaru, Einstein gravity as a nonholonomic almost K\"{a}%
hler geometry, Lagrange--Finsler variables, and deformation quantization, J. Geom. Phys. \textbf{60} (2010) 1289-1305

\bibitem{vqgr6} S. Vacaru, Two--connection renormalization and nonholonomic gauge models of Einstein gravity, Int. J. Geom. Meth. Mod. Phys. \textbf{7} (2010) 713-744

\bibitem{li} M. Li and Y. Pang, A trouble with Ho\v{r}ava--Lifshitz gravity, JHEP \textbf{0908} (2009) 015

\bibitem{sotir1} T. Sotiriou, M. Visser and S. Weinfurtner,
Phenomenologically viable Lorentz--violating quantum gravity, Phys. Rev. Lett. \textbf{102} (2009) 251601

\bibitem{sotir2} T. P. Sotiriou, M. Visser and S. Weinfurtner, Quantum
gravity without Lorentz invariance, JHEP \textbf{ 0910 } (2009) 033

\bibitem{bogdanos} C. Bogdanos and E. N. Saridakis, Perturbative
Instabilities in Ho\v{r}ava Gravity, Class. Quant. Grav. \textbf{ 27 } (2010) 075005

\bibitem{dimopoulos} S. Dimopoulos and G. Landsberg, Black holes at the  large hadron collider, Phys. Rev. Lett. \textbf{87} (2001) 161602

\bibitem{anchor} L. A. Anchordoqui, J. L. Feng, H. Goldberg and A. D.
Shapere, \ Black holes from cosmic rays: Probes of extra dimensions and new limits on TeV--scale gravity, Phys. Rev. D \textbf{65 } (2002) 124027

\bibitem{amelino} G. Amelino--Camelia, Gravity--wave interferometers as probes of a low--energy effective \ quantum gravity, Phys. Rev. D \textbf{62} (2000) 024015

\bibitem{midod} P. Midodashvili, Brane in 6D and localization of matter
fields, arXiv: hep-th/0308051

\bibitem{gm1} M. Gogberashvili and P. Midodashvili, Brane--universe in six dimensions, Phys. Lett. B \textbf{515} (2001) 447--450

\bibitem{gm2} M. Gogberashvili and P. Midodashvili, Localization of fields  on a brane in six dimensions, Europhys. Lett. \textbf{61} (2003) 308--313

\bibitem{gs1} M. Gogberashvili and D. Singleton, Nonsingular increasing gravitational potential for the brane in 5D, Phys. Lett. \textbf{B 582} (2004) 95--101

\bibitem{gs2} M. Gogberashvili and D. Singleton, Brane in 6D with increasing gravitational trapping potential, Phys. Rev. \textbf{D 69} (2004) 026004

\bibitem{singlbr} D. Singleton, Gravitational trapping potential with arbitrary extra dimensions, Phys. Rev. \textbf{D 70} (2004) 065013

\bibitem{ghodsi} A. Ghodsi and E. Hatefi, Extremal rotating solutions in Ho\v rava gravity, Phys. Rev. \textbf{ D 81 } (2010) 044016
\end{thebibliography}
\end{document}